%% file: HION-2016-07-PAPER.tex
\pdfoutput=1
\newcommand*{\ATLASLATEXPATH}{latex/}
\documentclass[cernpreprint, texlive=2016, UKenglish]{\ATLASLATEXPATH atlasdoc}

\usepackage{\ATLASLATEXPATH atlaspackage}
\usepackage{\ATLASLATEXPATH atlasbiblatex}

\usepackage[hion=true,process=true]{\ATLASLATEXPATH atlasphysics}
\usepackage{color,soul,verbatim}
\usepackage{tabularx}
\graphicspath{{logos/}{figures/}}

\input{HION-2016-07-PAPER-metadata}

\hypersetup{pdftitle={ATLAS document},pdfauthor={The ATLAS Collaboration}}

\begin{document}

\maketitle


\section{Introduction}
\label{sec:intro}
\input{Section1_Introduction}
\section{ATLAS detector}
\label{sec:detector}
\input{Section2_atlas-detector}
\section{Event and data selection}
\label{sec:data}
\input{Section3_Data-Selection}
\section{Data analysis}
\label{sec:Observables}
\input{Section4_Oberservable_determination}
\section{Systematic uncertainties}
\label{sec:Systematic}
\input{Section5_Systematics}
\section{Results}
\label{sec:result}
\input{Section6_Results}
 \FloatBarrier
\section{Summary}
\input{Section7_Conclusion}
\label{sec:conclusion}

\section*{Acknowledgements}

\input{acknowledgements/Acknowledgements}




\printbibliography
\clearpage
\input{atlas_authlist}



\end{document}

%% file: HION-2016-07-PAPER-metadata.tex
\AtlasTitle{Prompt and non-prompt $J/\psi$ and $\psi(2\mathrm{S})$ suppression at high transverse momentum in $5.02~\mathrm {TeV}$ Pb+Pb collisions with the ATLAS experiment}

\AtlasAbstract{%
  A measurement of $J/\psi$ and $\psi(2\mathrm{S})$ production is presented. It is based on a data sample from Pb+Pb collisions at  $\sqrt{s_{\mathrm{NN}}} = 5.02~\mathrm {TeV}$ and $pp$ collisions at $\sqrt{s} = 5.02~\mathrm {TeV}$ recorded by the ATLAS detector at the LHC in 2015, corresponding to an integrated luminosity of $0.42~\mathrm{nb}^{-1}$ and $25~\mathrm{pb}^{-1}$ in Pb+Pb and $pp$, respectively. The measurements of per-event yields, nuclear modification factors, and non-prompt fractions are performed in the dimuon decay channel for $9 < p_{T}^{\mu\mu} < 40$ GeV in dimuon transverse momentum, and $-2.0 < y_{\mu\mu} < 2.0$ in rapidity. Strong suppression is found in Pb+Pb collisions for both prompt and non-prompt $J/\psi$, as well as for prompt and non-prompt $\psi(2\mathrm{S})$, increasing with event centrality. The suppression of prompt $\psi(2\mathrm{S})$ is observed to be stronger than that of $J/\psi$, while the suppression of non-prompt $\psi(2\mathrm{S})$ is equal to that of the non-prompt $J/\psi$ within uncertainties, consistent with the expectation that both arise from \textit{b}-quarks propagating through the medium. 
Despite prompt and non-prompt $J/\psi$ arising from different mechanisms, the dependence of their nuclear modification factors on centrality is found to be quite similar.
}

\AtlasRefCode{HION-2016-07}

\PreprintIdNumber{CERN-EP-2018-049}

\AtlasJournal{Eur. Phys. J. C}

 \AtlasCoverSupportingNote{Supporting note for measurement of charmonia production with \PbPb data at $\sqrt{s_\textrm{NN}}=5.02$~\TeV}{https://cds.cern.ch/record/2196097/}
\AtlasCoverAnalysisTeam{Sebastian~Tapia, William~Brooks, Jorge~Lopez, Qipeng~Hu,  Nikita~Smirnov, Petr~Gallus}

\AtlasCoverEdBoardMember{Martin~Spousta~(chair), Andrew~Chisholm, Soumya~Mohapatra}

\AtlasCoverEgroupEditors{atlas-hion-2016-07-analysis-team@cern.ch}  

\AtlasCoverEgroupEdBoard{atlas-hion-2016-07-editorial-board@cern.ch}

%% file: Section1_Introduction.tex
Three decades ago, Matsui and Satz first suggested that charmonia, bound states of $c$- and $\bar{c}$-quarks, could be a sensitive probe to study the hot, dense system created in nucleus--nucleus (A+A) collisions~\cite{Matsui1986416}. They postulated that Debye screening of the quark colour charge in a hot plasma would lead to a dissociation of quarkonium bound state in the medium, such as \jpsi or $\psi(2\mathrm{S})$, when the Debye length becomes smaller than the quarkonium binding radius.
Therefore, the suppression of the quarkonia production should be significantly larger for $\psi(2\mathrm{S})$ than for \jpsi because the smaller binding energy facilitates the dissociation in the medium. This is referred to as sequential melting~\cite{Digal:2001ue,Karsch:2005nk}. In this picture, the suppression of different quarkonium states could therefore provide information related to the temperature and degree of deconfinement of the medium formed in heavy-ion collisions. 

There have been numerous experimental and theoretical investigations since then that have demonstrated that other effects are also present in addition to colour screening in a deconfined plasma~\cite{Lansberg:2008zm,Brambilla:2010cs,Andronic:2015wma}. First, it has been shown that over a wide range of interaction energies there is already a modification in the production of $\jpsi$ mesons in systems where a large volume of quark--gluon plasma does not appear to form, such as in proton--nucleus collisions~\cite{Ferreiro:2013pua,LHCB_pPb_816,CMF_phenix}. Second, it has been shown by the ALICE Collaboration that not only a suppression of quarkonia is observed in ion--ion collisions as reported by several collaborations~\cite{Aad:2010aa,CMS_1,PHENIX_1,Adamczyk:2013tvk}, but an enhancement is also observed at low transverse momentum, $\pT$, relative to production in proton--proton (\pp) collisions~\cite{ALICE_1,ALICE_2}. This observation has led to the interpretation that recombination of charm quarks and anti-quarks from the medium can play a role by providing an additional mechanism of quarkonium formation~\cite{Thews:2000rj,BraunMunzinger:2000px}. 

Finally, similarities between the suppression of \jpsi and the suppression of charged hadrons and $D$-mesons suggest that high-\pt $\jpsi$s may 
also be sensitive to parton energy loss in the medium~\cite{spousta,orleo}.
At LHC energies, $\jpsi$ originates not only from the immediate formation of the composite $ c \bar{c}$ bound state (prompt $\jpsi$), but also from the decay of \textit{b}-hadrons, which result in a decay vertex separated from the collision vertex by up to a few millimetres (non-prompt $\jpsi$). When a secondary vertex can be identified using the precise tracking system of the ATLAS experiment~\cite{PERF-2007-01}, it offers the intriguing possibility of using \jpsi production to study the propagation of \textit{b}-quarks in the hot dense medium, since hadronization is expected to take place outside the medium~\cite{Kopeliovich:2017imm}. Suppression of the production of \textit{b}-hadrons in the medium, in the most naive picture, is caused by a completely different phenomenon from the suppression of $c \bar{c}$ bound states. While $c\bar{c}$ bound state formation may be inhibited by colour screening from a hot and deconfined medium, the suppression of high-\pT \textit{b}-quark production is commonly attributed to energy loss of propagating \textit{b}-quarks by collisional or  radiative processes or both~\cite{Cao:2012au}, not necessarily suppressing the total cross section but more likely shifting the yield to a lower \pT. 
Quantum interference between the amplitudes for \textit{b}-hadron formation inside and outside of the nuclear medium may also play a role~\cite{Kopeliovich:2008uy}. 

The  modification of prompt $\jpsi$ production is not expected to be similar to the modification of non-prompt $\jpsi$ production, since quite different mechanisms can contribute to those two classes of final states~\cite{Andronic:2015wma}. Simultaneous measurements of prompt and non-prompt charmonia are therefore essential for understanding the physics mechanisms of charmonium suppression in heavy-ion collisions.

This paper reports measurements of prompt and non-prompt per-event yields, non-prompt fraction and nuclear modification factors, $R_{\mathrm{AA}}$, of the $\Jpsi$ and $\psi(2\mathrm{S})$. The measurements are performed in the dimuon decay channel in $\PbPb$ collisions at $\sqrt{s_{\mathrm{NN}}}$ = 5.02 TeV, with an integrated luminosity of $0.42~\mathrm{nb^{-1}}$, and in $\pp$ collisions at $\sqrt{s}$ = 5.02 TeV, with an integrated luminosity of $25~\ipb$. The data were recorded during the 2015 LHC $\PbPb$ and $\pp$ runs. The measurements are presented for $9 < \pt^{\mu\mu} < 40\ \GeV$ in dimuon transverse momentum, and $-2.0 < y_{\mu\mu} < 2.0$ in rapidity. 

Section~2 describes the ATLAS detector, Section~3 discusses the selection procedure applied to  the data, the data analysis is presented in Section~4 and systematic
uncertainties in Section~5. Results and a summary of the paper are presented in Section 6 and 7.

%% file: Section2_atlas-detector.tex
\newcommand{\AtlasCoordFootnote}{%
ATLAS uses a right-handed coordinate system with its origin at the nominal interaction point (IP)
in the centre of the detector and the $z$-axis along the beam pipe.
The $x$-axis points from the IP to the centre of the LHC ring,
and the $y$-axis points upwards.
Cylindrical coordinates $(r,\phi)$ are used in the transverse plane, 
$\phi$ being the azimuthal angle around the $z$-axis.
The pseudorapidity is defined in terms of the polar angle $\theta$ as $\eta = -\ln \tan(\theta/2)$.
Angular distance is measured in units of $\Delta R \equiv \sqrt{(\Delta\eta)^{2} + (\Delta\phi)^{2}}$.}

The ATLAS detector~\cite{PERF-2007-01} at the LHC covers nearly the entire solid angle around the collision point.\footnote{\AtlasCoordFootnote}
It consists of an inner tracking detector surrounded by a thin superconducting solenoid, electromagnetic and hadronic calorimeters,
and a muon spectrometer incorporating three large superconducting air-core toroid magnets with eight coils each.

The inner-detector system is immersed in a \SI{2}{\tesla} axial magnetic field 
and provides charged-particle tracking in the pseudorapidity range $|\eta| < 2.5$.
A high-granularity silicon pixel detector covers the vertex region and typically provides three measurements per track, 
the first hit being normally in the innermost layer.
Since 2015 the detector has been augmented by the insertable B-layer~\cite{Capeans:2010jnh}, an additional pixel layer close to the interaction point which provides high-resolution hits at small radius to improve the tracking and vertex reconstruction performance, significantly contributing to the reconstruction of displaced vertices.
It is followed by a silicon microstrip tracker which comprises eight cylindrical layers of single-sided silicon strip detectors in the barrel region, and nine disks in the endcap region. 
These silicon detectors are complemented by a transition radiation tracker (TRT),
which enables radially extended track reconstruction up to $|\eta| = 2.0$.

The calorimeter system covers the pseudorapidity range $|\eta| < 4.9$. Within the region $|\eta|< 3.2$, electromagnetic calorimetry is provided by barrel and endcap high-granularity lead/liquid-argon (LAr) calorimeters, with an additional thin LAr presampler covering $|\eta| < 1.8$, to correct for energy loss in material upstream of the calorimeters. Hadronic calorimetry is provided by a steel/scintillator-tile calorimeter, segmented into three barrel structures within $|\eta| < 1.7$, and two copper/LAr hadronic endcap calorimeters situated at $1.5 < |\eta| < 3.2$. The solid angle coverage is completed with forward copper/LAr and tungsten/LAr calorimeter modules (FCal) situated at $3.1 < |\eta| < 4.9$, optimized for electromagnetic and hadronic measurements respectively.

The muon spectrometer comprises separate trigger and high-precision tracking chambers measuring the deflection of muons in a magnetic field generated by the superconducting air-core toroids. The precision chamber system covers the region $|\eta| < 2.7$ with three layers of monitored drift tubes, complemented by cathode strip chambers in the forward region, where the background is the highest. The muon trigger system covers the range of $|\eta| < 2.4$ with resistive plate chambers in the barrel, and thin gap chambers in the endcap regions.

In addition to the muon trigger, two triggers are used in Pb+Pb collisions to select minimum-bias events for the centrality analysis. These are based on the presence of a minimum amount of transverse energy in all sections of the calorimeter system ($|\eta| < 3.2$) or, for events which do not meet this condition, on the presence of substantial energy deposits in both zero-degree calorimeters (ZDCs), which are primarily sensitive to spectator neutrons in the region $|\eta| > 8.3$. Those two triggers were found to be fully efficient in the centrality range studied in this analysis.

A two-level trigger system is used to select events of interest~\cite{Aaboud:2016leb}. The first-level (L1) trigger is implemented in hardware and uses a subset of detector information to reduce the event rate to a design value of at most \SI{100}{\kHz}. This is followed by a software-based high-level trigger (HLT), which reduces the event rate to a maximum value of \SI{1}{\kHz}.

%% file: Section3_Data-Selection.tex
The analysis presented in this paper uses data from \PbPb collisions at a nucleon--nucleon centre-of-mass energy of \sqn = 5.02 TeV and \pp collisions at a centre-of-mass energy of $\sqrt{s}$ = 5.02 TeV recorded by the ATLAS experiment in 2015. The integrated luminosity of the analysed samples is $0.42~\mathrm{nb^{-1}}$ for \PbPb collisions and $25~\ipb$ for \pp collisions.

Events were collected using a trigger requiring that the event contains at least two reconstructed muons. In the $\pp$ sample both muons must generate a L1 muon trigger and be confirmed by the HLT while in the \PbPb sample only one muon is required to be seen by the L1 muon trigger and confirmed by the HLT; the second muon is only required to pass the HLT. At both levels the muon must satisfy the requirement of $p_\mathrm{T} > 4$~GeV, as reconstructed by the trigger system.

Monte Carlo (MC) simulations are used for performance studies, where the response of the ATLAS detector was simulated using Geant 4~\cite{SOFT-2010-01,AGOSTINELLI2003250}.  Prompt ($\pp\rightarrow\jpsi\rightarrow\mu\mu$) and non-prompt
($\pp\rightarrow\bbbar\rightarrow\jpsi\rightarrow\mu\mu$) samples of $\jpsi$ were produced with the event generator \textsc{Pythia 8.212}~\cite{Sjostrand:2014zea} and corrected for electromagnetic radiation with \textsc{Photos}~\cite{PHOTOS}. The A14 set of tuned parameters~\cite{ATL-PHYS-PUB-2014-021} is used together with the CTEQ6L1 parton distribution function set~\cite{Pumplin:2002vw}. These samples are used to study the trigger and reconstruction performance of the $\pp$ collisions. In order to simulate \jpsi production in the high multiplicity environment of \PbPb collisions, the generated events were overlaid with a sample of multi-jet events produced with \textsc{HIJING}~\cite{HIJING}.  

Muon candidates are required to pass the "tight" muon working point selection~\cite{MuPerf:2015} without any TRT requirements, have $\pt>4$ GeV, and $|\eta|<2.4$ in addition to being consistent with the reconstructed muon associated with the trigger decision.
To be selected, a muon pair must be consistent with originating from a common vertex, have opposite charge, and an invariant mass in the range $2.6<m_{\mu\mu}<4.2$ GeV. The dimuon candidate is further required to have $p_\textrm{T}^{\mu\mu} > 9$ GeV to ensure that the pair candidates are reconstructed in a fiducial region where acceptance and efficiency corrections 
do not vary significantly relative to the systematic uncertainties in the final results.

The centrality of \PbPb collisions is characterized by the sum of the transverse energy, $\sum E_\textrm{T}^\mathrm{FCal}$, evaluated at the electromagnetic scale in the FCal. It describes the degree of geometric overlap of two colliding nuclei in the plane perpendicular to the beam with large overlap in central collisions and small overlap in peripheral collisions. Centrality intervals are defined in successive percentiles of the $\sum E_\textrm{T}^\mathrm{FCal}$ distribution ordered from the most central (highest $\sum E_\textrm{T}^\mathrm{FCal}$ ) to the most peripheral collisions. A Glauber model analysis of the $\sum E_\textrm{T}^\mathrm{FCal}$ distribution was used to evaluate the mean nuclear thickness function, \avgTAA, and the number of nucleons participating in the collision, \avgNpart, in each centrality interval~\cite{Steinberg:0701025,Steinberg:0805.4411,CONF.123}. The centrality intervals used in this measurement are indicated in Table \ref{tab:centrality} along with their respective calculations of \avgTAA and \avgNpart.

\begin{table}[t]
   \centering
   \begin{tabular}{@{} crr @{}} 
      \toprule
      Centrality [\%]    & \avgTAA [$\mathrm{mb}^{-1}$]  & \avgNpart \\
      \midrule
      0--5    & 26.23 $\pm$ 0.22 & 384.4 $\pm$ 1.9 \\
      5--10    & 20.47 $\pm$ 0.19 & 333.1 $\pm$ 2.7 \\
      0--10    & 23.35 $\pm$ 0.20 & 358.8 $\pm$ 2.3\\
      10--20  & 14.33 $\pm$ 0.17 & 264.0 $\pm$ 2.8 \\
      20--30  & 8.63 $\pm$ 0.17 & 189.1 $\pm$ 2.7   \\
      30--40  & 4.94 $\pm$ 0.15 & 131.4 $\pm$ 2.6  \\
      40--50  & 2.63 $\pm$ 0.11 & 87.0 $\pm$ 2.3  \\
      50--60  & 1.27 $\pm$ 0.07 & 53.9 $\pm$ 1.9   \\
      60--80  & 0.39 $\pm$ 0.03 & 22.9 $\pm$ 1.2 \\
      20--50  & 5.40 $\pm$ 0.14 & 135.8 $\pm$ 2.5 \\
      0--80  & 6.99 $\pm$ 0.10 & 141.3 $\pm$ 2.0  \\
      \bottomrule
   \end{tabular}
   \caption{The \avgTAA, \avgNpart values and uncertainties in each centrality bin. These are the results from the Glauber modelling of the summed transverse energy in the forward calorimeters, $\sum E_\textrm{T}^\mathrm{FCal}$.}
   \label{tab:centrality}
\end{table}

The number of minimum-bias events, $N_{\textrm{evt}}$, is used to normalize the yield in respective centrality class. Minimum-bias events are selected by requiring that they pass at least one of the two minimum-bias triggers. The analysed dataset corresponds to $2.99 \times 10^{9}$ \PbPb collisions after correction for the trigger prescale factor. 

%% file: Section4_Oberservable_determination.tex
The pseudo-proper decay time, $\tau$, is used to distinguish between prompt and non-prompt charmonium production.
It is defined as,
\begin{equation*}
        \tau = \frac{L_{xy}m_{\mu\mu}}{\pt^{\mu\mu}},
\label{eq:pseudo_proper_time_def}
\end{equation*}
where $L_{xy}$ is the distance between the position of the reconstructed dimuon vertex and the primary vertex projected onto the transverse plane.
A weight, $w_{\mathrm{total}}$, is defined for each selected dimuon candidate using the relation:
\begin{equation*}
w^{-1}_{\mathrm{total}} = A\times\epsilon_{\mathrm{reco}}\times\epsilon_{\mathrm{trig}},
\label{eq:weight}
\end{equation*} 
where $A$ is the acceptance, $\epsilon_{\mathrm{reco}}$ is the reconstruction efficiency, and $\epsilon_{\mathrm{trig}}$ is the trigger efficiency. 

A two-dimensional unbinned maximum-likelihood fit to the invariant mass and pseudo-proper time distributions of weighted events is used to determine 
the yields of the prompt and non-prompt charmonium components as well as the contribution from background. A total of 31\ 572 events before applying the weights
are used in the fit. 

The differential cross sections for the production of prompt (p) and non-prompt (np) \jpsi and $\psi(2\mathrm{S})$ in \pp collisions were calculated
in a previously published study~\cite{Qipeng:PaperpPbJpiATLAS} and are defined as:
\begin{equation*}
        \frac{\text{d}^2\sigma^\text{p(np)}}{\text{d}\pt \text{d}y}\times B(\psi(\textit{n}\mathrm{S})\rightarrow\mu\mu) = \frac{N_{\psi(\textit{n}\mathrm{S})}^\mathrm{p(np),~corr}}{\Delta\pt\times\Delta y\times \int{\mathcal{L}dt}},
\label{eq:prompt_xsection_def}
\end{equation*}
where $B(\psi(\textit{n}\mathrm{S})\to\mu\mu)$ is the branching ratio for charmonium states decaying into two muons~\cite{Agashe:2014kda}, $N_{\psi(\textit{n}\mathrm{S})}^\mathrm{p(np),~corr}$ 
is the prompt and non-prompt charmonium yield corrected for acceptance and detector effects, and $\Delta\pt$ and $\Delta y$ are the widths of the \pt and $y$ bins. 
Following the same approach,  the per-event yield of charmonium states measured in A+A collisions is calculated as:
\begin{equation}
       \left. \frac{\text{d}^{2}N^\text{p(np)}}{\text{d}\pt \text{d}y}\right|_\textrm{cent} \times B(\psi(\textit{n}\mathrm{S})\rightarrow\mu\mu) = \frac{1}{\Delta\pt\times\Delta y}\times \left.\frac{N_{\psi(\textit{n}\mathrm{S})}^\mathrm{p(np),~corr}}{N_\textrm{evt}}\right|_\textrm{cent},
\label{eq:prompt_perEventYield_def}
\end{equation}
where $N_\textrm{evt}$ is the number of minimum-bias events and "cent" refers to a specific centrality class.

\subsection{Acceptance and efficiency corrections}
\label{sec_corr}
The kinematic acceptance $A(\pt,y)$ for a $\psi(\textit{n}\mathrm{S})$ with transverse momentum $\pt$ and rapidity $y$ decaying into $\mu\mu$ was 
obtained from a MC simulation and is defined as the probability that both muons fall within the fiducial volume $p_\textrm{T}(\mu^{\pm})>4$ GeV and $|\eta(\mu^{\pm})|<2.4$. 
Acceptance generally depends on the $\psi(\textit{n}\mathrm{S})$ polarization. In this study, we assume that the $\psi(\textit{n}\mathrm{S})$ are unpolarized following Refs.~\cite{LHCb:Polarization,Alice:Polarization,CMS:Polarization}. The effects of variations to this assumption have been considered and are discussed in Section~\ref{sec:Systematic}.
In order to apply the acceptance correction to the yields, a simple linear interpolation is used assuming the upper mass boundary for the $J/\psi$ candidates to be 3.5~GeV and the lower 
mass boundary for the $\psi(2\mathrm{S})$ candidates to be 3.2~GeV, resulting in a full range of 0.3 GeV. Within the interpolation range 
of $m_{\mu\mu}$~=~3.2--3.5~GeV, the following function was applied for the acceptance correction:
\begin{equation}
        A =A(J/\psi) \times  \frac {3.5 - m_{\mu\mu}}{0.3} +
        A(\psi(2\mathrm{S})) \times \frac {m_{\mu\mu} - 3.2}{0.3}.
\label{eq:A_def}
\end{equation}

Trigger and reconstruction efficiencies were calculated for both data and MC simulation using the tag-and-probe (T\&P) method. The method is based on the selection of an almost pure muon sample from $J/\psi \rightarrow \mu\mu$ events collected with an auxiliary single-muon trigger, requiring one muon of the decay (tag) to be identified as the "tight" muon which triggered the read-out of the event and the second muon (probe) to be reconstructed as a system independent of the one being studied, allowing a measurement of the performance with minimal bias. Once the tag and probe sample is defined, the background contamination and the muon efficiency are measured with a simultaneous maximum-likelihood fit of two statistically independent distributions of the invariant mass: events in which the probe is or is not successfully matched to the selected muon~\cite{MuPerf:2015,BookDataAnalysisHEP}. 
Both efficiencies were evaluated as a function of \pT and $\eta$ using muons from simulated $J/\psi\rightarrow\mu\mu$ decays. 
Muon reconstruction efficiency increases from low to high \pT and decreases from central to forward rapidities. It varies between 60\% and 90\%, becoming almost constant for $\pt>6$ GeV. 
The dimuon trigger efficiency is studied and factorized in terms of single-muon trigger efficiencies which increase from low to high \pT and from central to forward rapidities. Dimuon trigger efficiency increases from 50\% to 85\% between the lowest and highest dimuon \pT.

\begin{figure}[htb]
  \begin{center}
  \includegraphics[width=0.49\textwidth]{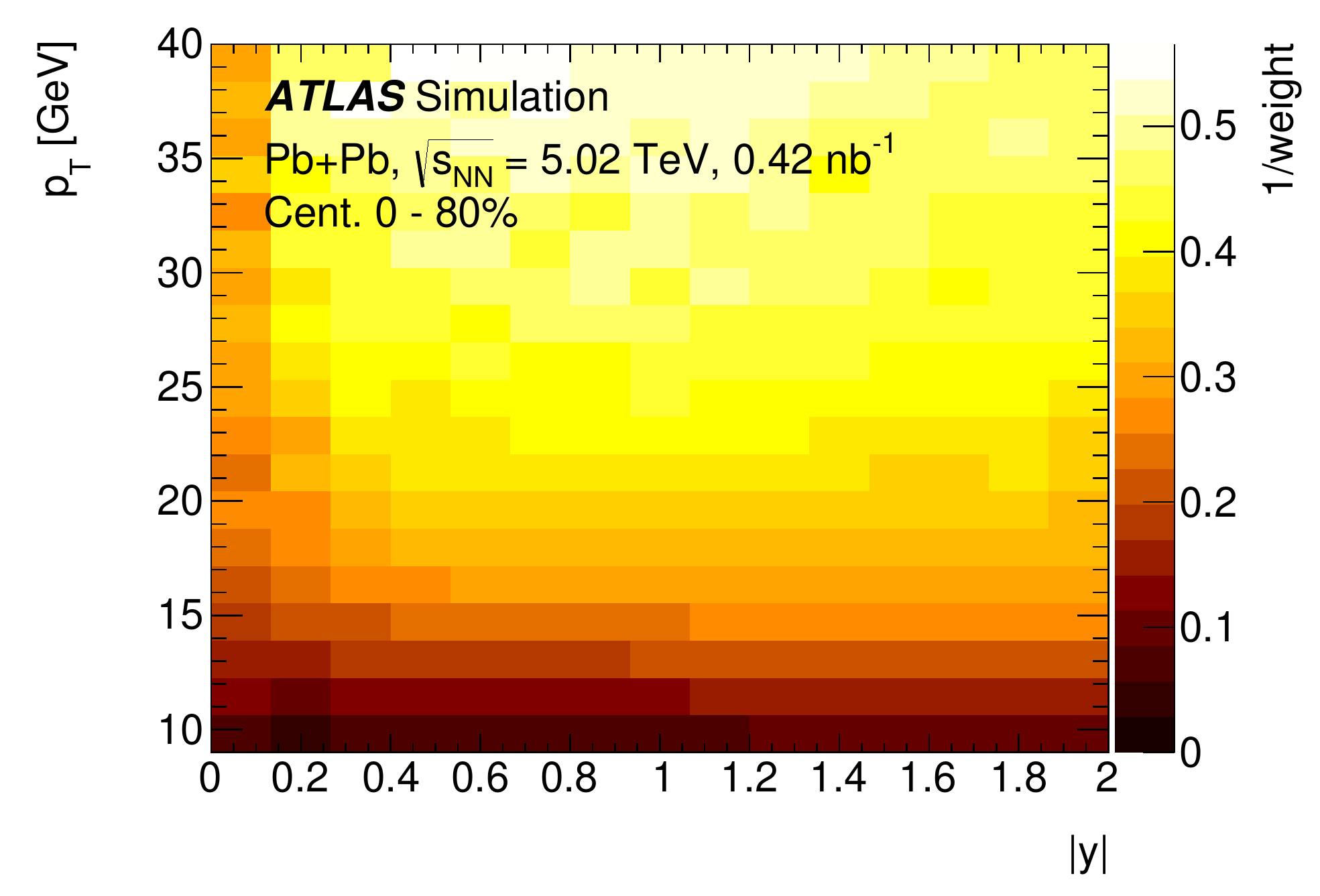}
   \includegraphics[width=0.49\textwidth]{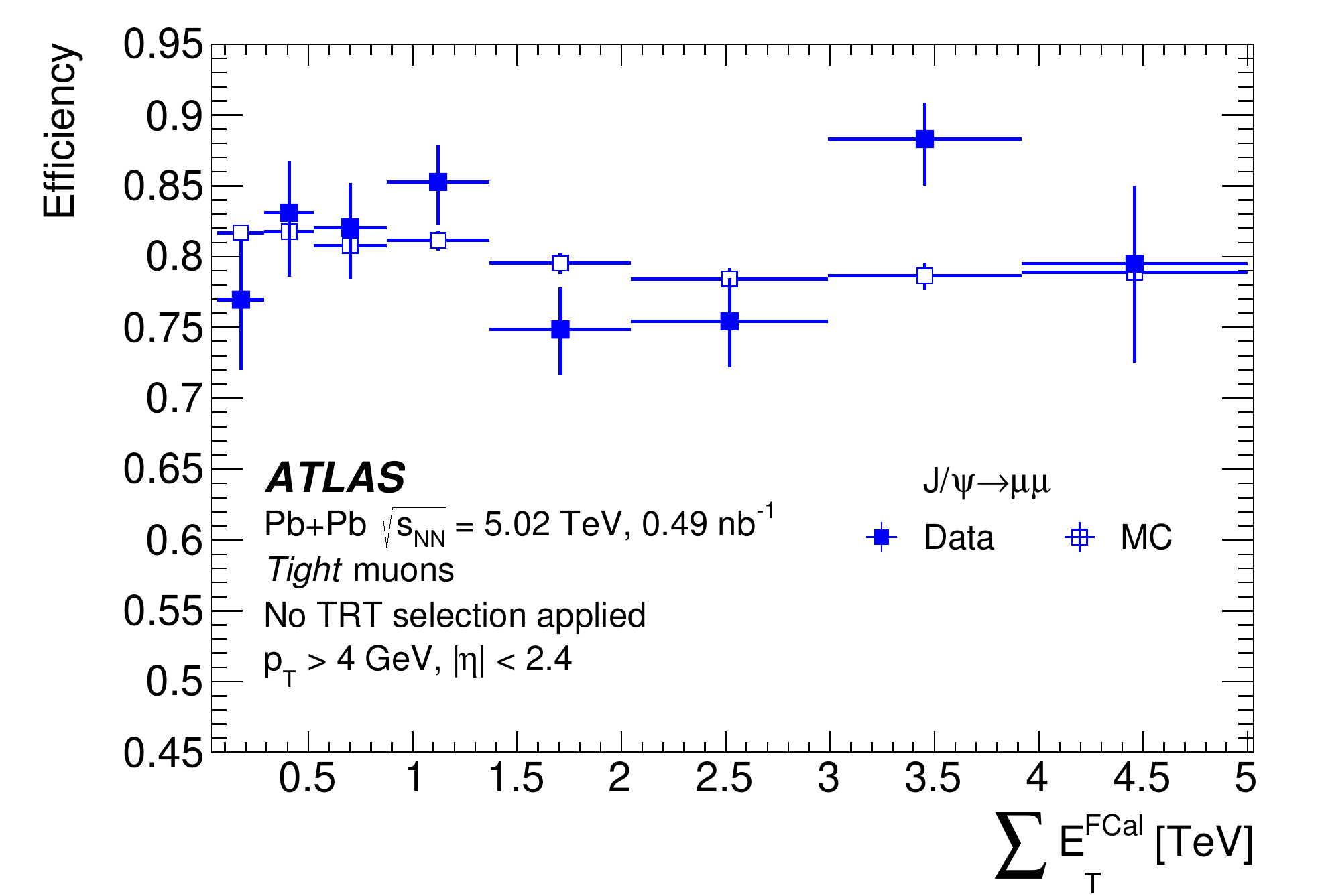}
   \caption{(Left) Inverse total weight binned in the dimuon transverse momentum and rapidity for integrated centrality as
   estimated in MC simulation and corrected for differences between efficiencies in MC and experimental data. Decreases in efficiency at very central rapidity correspond to the $|\eta|<0.1$ region not covered by the muon detectors. The weight is dominated by the acceptance correction. (Right) Muon reconstruction efficiency as a function of the summed transverse energy in the forward calorimeters, $\sum E_\textrm{T}^\mathrm{FCal}$.}
  \label{fig:total_w}
  \end{center}
\end{figure}

In order to account for the difference between efficiencies in simulation and experimental data, the 
data-to-MC ratio, $\epsilon^\mathrm{data}_{\mathrm{reco}}/\epsilon^{\mathrm{MC}}_{\mathrm{reco}}$, was parameterized as a 
function of $p_\textrm{T}$ and centrality and applied as a multiplicative scale factor to the efficiency correction separately for the barrel and endcap regions of the muon spectrometer. This scale factor varies between 1.01 to 1.05.
The inverse total weight, $w^{-1}_{\mathrm{total}}$, after applying the scale factor, is shown in the left panel of Figure~\ref{fig:total_w}, averaged in bins 
of the dimuon transverse momentum and rapidity. The right panel of Figure~\ref{fig:total_w} shows the centrality dependence of the muon reconstruction efficiency.
\subsection{Fit model}
\label{Sec41}
The corrected prompt and non-prompt $\psi(\textit{n}\mathrm{S})$ yields are extracted from two-dimensional weighted unbinned maximum-likelihood fits performed on invariant mass and pseudo-proper decay time distributions. A separate fit is made for the yields and the nuclear modification factor for each \pt, $y$, and 
centrality interval that is measured in this analysis.
The probability distribution function (PDF) for the fit~\cite{BPHY-2012-02} is defined as a normalized sum of seven terms listed in Table~\ref{tab:pdf}, where each term is factorized into mass-dependent and decay-time-dependent functions; these functions are described below. The PDF can be written in a compact form as:
\begin{equation*}
	\text{PDF}(m,\tau) = \sum^7_{i=1}\kappa_i f_i(m) \cdot h_i(\tau) \otimes g(\tau),
\label{eq:pdfdef}
\end{equation*}
where $\kappa_i$ is the normalization factor of each component, $f_i(m)$ and $h_i(\tau)$ are distribution functions for the mass $m$ and the pseudo-proper time $\tau$ respectively; $g(\tau)$ is the resolution function described with a double Gaussian distribution; and the "$\otimes$" symbol denotes a convolution. The distribution functions $f_i$ and $h_i$ are defined by a Crystal Ball ($C\!B$) function~\cite{paper:CB}, Gaussian ($G$), Dirac delta ($\delta$) and exponential ($E$) distributions; individual components are shown in Table~\ref{tab:pdf}. 
\begin{table}[t]
\begin{center}
\begin{tabular}{ccccc}
  \hline
  $i$ & Type           & Source    & $f_i(m)$                                  				& $h_i(\tau)$ \\ \hline
  1 & $\jpsi$        & p         & $\omega\ C\!B_1(m)+(1-\omega)G_1(m)$  & $\delta(\tau)$ \\
  2 & $\jpsi$        & np        & $\omega\ C\!B_1(m)+(1-\omega)G_1(m)$  & $E_1(\tau)$ \\
  3 & $\psi(2\mathrm{S})$     & p         & $\omega\ C\!B_2(m)+(1-\omega)G_2(m)$  & $\delta(\tau)$ \\
  4 & $\psi(2\mathrm{S})$     & np        & $\omega\ C\!B_2(m)+(1-\omega)G_2(m)$  & $E_2(\tau)$ \\ \hline
  5 & Bkg      		 & p         & $E_3(m)$                                     		& $\delta(\tau)$ \\
  6 & Bkg      		 & np        & $E_4(m)$                                  		& $E_5(\tau)$ \\
  7 & Bkg      		 & np        & $E_6(m)$                                  		& $E_7(|\tau|)$ \\
  \hline
\end{tabular}
\caption{Probability distribution functions for individual components in the default fit model used to extract the prompt (p) and non-prompt (np) contribution for $\jpsi$ and $\psi(2\mathrm{S})$ signal and background (Bkg). 
Symbols denote functions as follows: "$C\!B$" -- Crystal Ball, "$G$" -- Gaussian, "$E$" -- exponential, and "$\delta$" -- Dirac delta function.}
\label{tab:pdf}
\end{center}
\end{table}
The fit is performed using the RooFit framework~\cite{Roofit:2003}. In order to stabilize the fit model, and reduce the correlation between parameters, a number of component terms listed in Table 2 share common parameters, are scaled to each other by a multiplicative scaling parameter, or are fixed to the value observed in MC simulation.

The signal mass shapes of the \jpsi and $\psi(2\mathrm{S})$ are each described by the sum of a $C\!B$ function, which covers the \jpsi invariant mass distribution's low-side tail due to final-state radiation, and a single Gaussian function which share a common peak position. The width term in the $C\!B$ function is equal to the Gaussian standard deviation times a free scaling term that is common to the $\jpsi$ and $\psi(2\mathrm{S})$. The $C\!B$ low-mass tail and height parameters are fixed. Variations of these two parameters are considered a part of the fit model's systematic uncertainties. The mean of the $\psi(2\mathrm{S})$ mass profile is set to be the mean of the \jpsi mass profile multiplied by the ratio of their known masses, $m_{\psi(2\mathrm{S})}/m_{\jpsi} = 1.190$~\cite{Agashe:2014kda}.
The Gaussian width of the $\psi(2\mathrm{S})$ is also set to be the width of the $\jpsi$ multiplied by the same factor. Variations of this scaling term are considered a part of the fit model systematic uncertainties. The relative fraction of the $C\!B$ and Gaussian functions, $\omega$, is free but common to the $\jpsi$ and $\psi(2\mathrm{S})$.

The non-prompt signal pseudo-proper decay time PDFs are described by a single-sided exponential function (for positive $\tau$ only) convolved with a double Gaussian lifetime resolution function. The double Gaussian resolution function has a fixed mean at $\tau = 0$ and free widths with a fixed relative fraction for the two single Gaussian components. The same resolution function is used to describe the prompt contribution by convolving it with a delta function.                                            

The pseudo-proper decay time PDFs describing the background are represented by the sum of one prompt component and two non-prompt components. 
The prompt background component is described by a delta function convolved with a double Gaussian function. 
While one of the non-prompt background contributions is described by a single-sided decay model (for positive $\tau$ only), the other is described 
by a double-sided decay model accounting for candidates of mis-reconstructed or non-coherent dimuon pairs resulting from other Drell--Yan muons and combinatorial background. 
The same Gaussian resolution functions are used for the background and the signal. For the background parameterizations in the mass distribution, the 
three components: prompt, single-sided non-prompt, and double-sided non-prompt were modelled with exponentials functions.

\begin{figure}[htb]
  \begin{center}
  \includegraphics[width=0.49\textwidth]{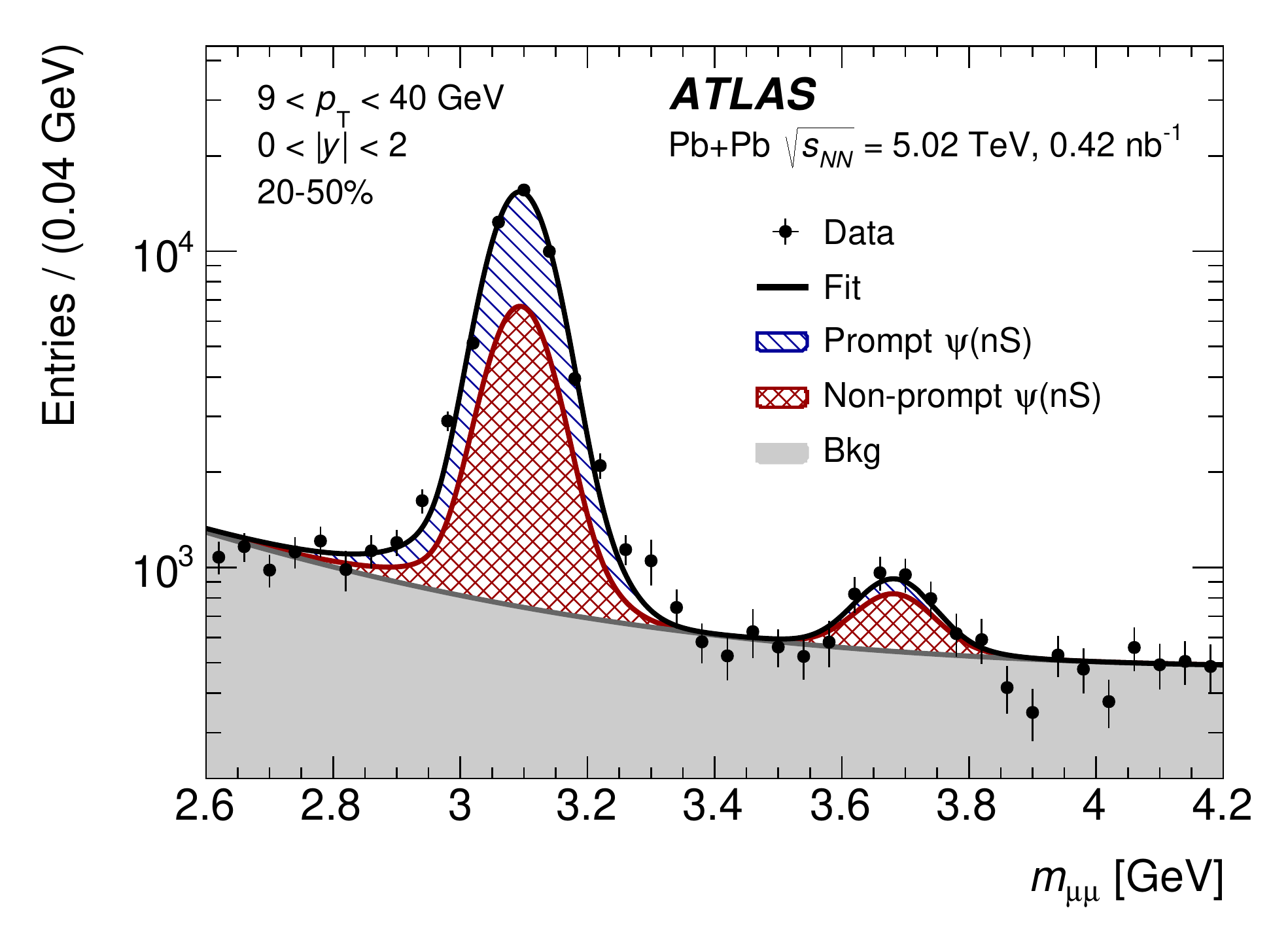}
  \includegraphics[width=0.49\textwidth]{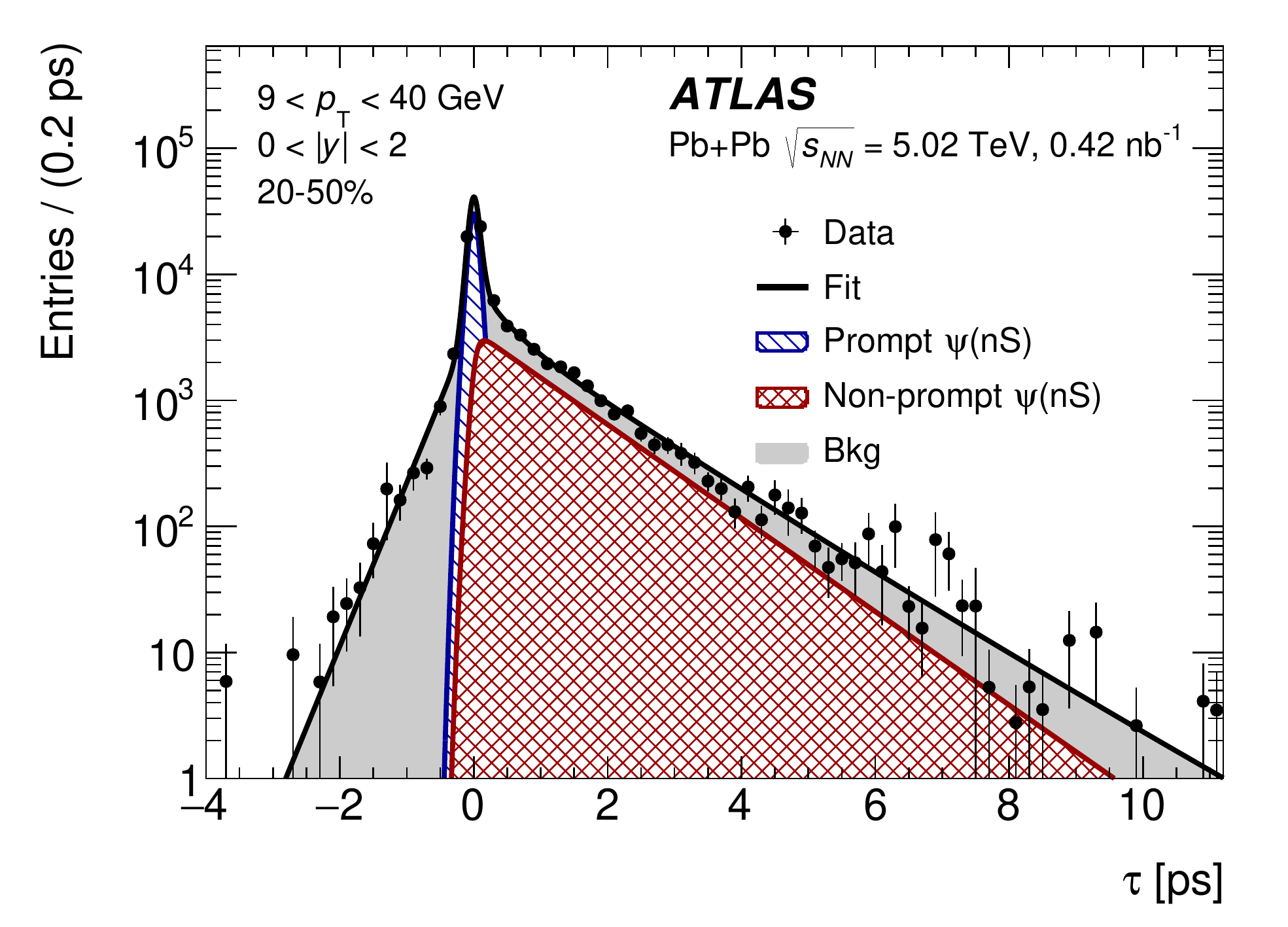}\\
  \caption{Dimuon invariant mass for events with $2.6 < m_{\mu\mu} < 4.2\ \GeV$ (left) and dimuon pseudo-proper lifetime (right). The data, corrected for acceptance times efficiency, are shown for the range $ 9 < \pt < 40\ \GeV$, $|y| < 2.0$, and centrality 20--50\% in \PbPb collisions. Superimposed on the data are the projections of the fit results.}
  \label{fig:2D_fit_PbPb}
  \end{center}
\end{figure}

Example fit projections are shown in Figure~\ref{fig:2D_fit_PbPb}. The important quantities extracted from the fit are: the number of signal $J/\psi$, the number of signal $\psi(2\mathrm{S})$, the non-prompt fraction of the $J/\psi$ signal, and the non-prompt fraction of the $\psi(2\mathrm{S})$ signal. From these values and the correlation matrix of the fit, all measured values are extracted.

\subsection{Observables}
The suppression of charmonium states is quantified by the nuclear modification factor, which can be defined for a given centrality class as:
\begin{equation}
      R_{\mathrm{AA}} = \frac{N_{\mathrm{AA}}}{\langle T_{\mathrm{AA}}\rangle \times\ \sigma_{\pp}},
\label{eq:RPbPb_def2}
\end{equation}
where $N_\mathrm{AA}$ is the per-event yield of charmonium states measured in A+A collisions, $\langle T_\mathrm{AA}\rangle$ is the mean nuclear thickness function and $\sigma_{\pp}$ is the cross section for the production of the corresponding charmonium states in $\pp$ collisions at the same energy \cite{Qipeng:PaperpPbJpiATLAS}. 

In order to quantify the production of $\psi(\mathrm{2\mathrm{S}})$ relative to \jpsi a ratio of nuclear modification factors, $\rho_\textrm{PbPb}^{\psi(2\mathrm{S})/\jpsi}$ = $R_\mathrm{AA}^{\psi(\mathrm{2\mathrm{S}})}/R_\mathrm{AA}^{\jpsi}$, can be used. However, in this analysis the numerator and denominator are not calculated directly from Eq.~(\ref{eq:RPbPb_def2}), rather, it is advantageous to calculate it in the equivalent form as:
\begin{equation*}
\rho_\textrm{PbPb}^{\psi(2\mathrm{S})/\jpsi} = ( N_{\psi(2\mathrm{S})} / N_{\jpsi} )_{\PbPb} /( N_{\psi(2\mathrm{S})} / N_{\jpsi} )_{\pp}.
\end{equation*}
This formulation minimizes the systematic uncertainties due to a substantial cancelling-out of the trigger and reconstruction efficiencies for the two quarkonia systems because they are very similar in mass and they are measured in the identical final-state channel.

Also measured is the non-prompt fraction $f_\mathrm{np}$, which is defined as the ratio of the number of non-prompt charmonia to the number of inclusively produced charmonia,
\begin{equation*}
        f_\mathrm{np}^{\psi(\textit{n}\mathrm{S})} = \frac{N_{\psi(\textit{n}\mathrm{S})}^\mathrm{np,corr}}{N_{\psi(\textit{n}\mathrm{S})}^\mathrm{np,corr} + N_{\psi(\textit{n}\mathrm{S})}^\mathrm{p,corr}},
\label{eq:nonprompt_fraction_def}
\end{equation*}
where the non-prompt fraction can be determined for the $\jpsi$ and $\psi(2\mathrm{S})$ simultaneously. This observable has the advantage that acceptances and efficiencies are similar for the numerator and denominator, and thus systematic uncertainties are reduced in the ratio.

%% file: Section5_Systematics.tex
The main sources of systematic uncertainty in this measurement are the assumptions in the fitting procedure, the acceptance and efficiency calculations, and the $pp$ luminosity and $\langle T_\mathrm{AA}\rangle$ determination.
The acceptance, and hence the corrected yields, depend on the spin-alignment state of the $\psi(\textit{n}\mathrm{S})$. For prompt production, six alternative scenarios have been considered, corresponding to extreme cases of spin alignment, as explained in Ref.~\cite{BPHY-2012-02}. An envelope to the acceptance has been obtained from the maximum deviations from the assumption of unpolarized production.
In the non-prompt case a map weighted to the CDF result~\cite{Abulencia:2007us} for $B\rightarrow\jpsi$ spin-alignment is used as a variation. Since the polarization of charmonia in $pp$ collisions was measured to be small~\cite{LHCb:Polarization,Alice:Polarization,CMS:Polarization}, its modification due to the nuclear environment is neglected and the spin-alignment uncertainty is assumed to cancel out in \RAA and $\rho_\textrm{PbPb}^{\psi(2\mathrm{S})/\jpsi}$. Changes in the yields due to bin migration effects are at the per-mil level and thus no correction is needed.
Table~\ref{tab:syst} shows the systematic uncertainties affecting the three measured observables. The total systematic uncertainty is calculated by summing the different contributions in quadrature and is derived separately for $pp$ and Pb+Pb results. The yield extraction uncertainties increase from central to forward rapidity, and from high to low \pT, where the dominant source of systematic uncertainty is the muon reconstruction. The double $R_\mathrm{AA}$ ratio, $\rho_\textrm{PbPb}^{\psi(2\mathrm{S})/\jpsi}$ has a substantially larger fit uncertainty than the other observables; this is because the signal-to-background ratio for the $\psi(2\mathrm{S})$ is much smaller than for the $\Jpsi$. For those two observables the correlations between the uncertainty in the $pp$ and \PbPb samples are taken into account.
\begin{table}[t]
   \centering
  \begin{tabularx}{0.8\textwidth}{@{}Xcrcrrc@{}}
\toprule
 		& \multicolumn{2}{c}{\jpsi yield} & \multicolumn{2}{c}{$R^{J/\psi}_\textrm{AA}$} & \multicolumn{1}{c}{$\rho_\textrm{PbPb}^{\psi(2\mathrm{S})/\jpsi}$}\\
Source 	& Uncorr.    & Corr. & Uncorr. & Corr. & Uncorr.  \\
\midrule
 Trigger                  & 2 - 4\%     & 3\%         & 5 - 6\%     & 5\% & $<1\%$       \\
Reconstruction      & 4 - 5\%   & 2\%         & 6 - 7\%   & 2\%  & $<1\%$    \\
Fitting                    & 1 - 2\%     & 1\%         & 1 - 2\%     & 1\%  & 8 - 9\%    \\
\TAA                    & --     & 1 - 8\%         & --     & 1 - 8\%  & --    \\
Luminosity             & --     & --         & --     & 5.4\%  & --    \\
\bottomrule
\end{tabularx}  
\caption{Systematic uncertainties of the $\jpsi$ yield, $R^{J/\psi}_\textrm{AA}$ and $\rho_\textrm{PbPb}^{\psi(2\mathrm{S})/\jpsi}$ measured in \PbPb collisions. "Uncorr." refers to point-to-point uncorrelated uncertainties and "Corr." refers to global uncertainties from various sources.}
\label{tab:syst}
\end{table}

\subsection{Proton--proton luminosity and mean nuclear thickness uncertainties}
The integrated luminosity determined for the 2015 \pp data was calibrated using data from dedicated beam-separation scans, also known as van der Meer scans. Sources of systematic uncertainty similar to those examined in the 2012 $\pp$ luminosity calibration \cite{ATLAS:Lumi} were studied in order to assess the systematic uncertainties for the 2015 data. The combination of these systematic uncertainties results in a uncertainty in the luminosity during \pp collisions at $\sqrt{s} = 5.02\ \TeV$ of $\delta \mathcal{L}/\mathcal{L} = \pm 5.4\%$. 
The uncertainty in the value of the nuclear overlap function \avgTAA is estimated by varying the Glauber model parameters and is shown in Table~\ref{tab:centrality}. This uncertainty is treated as fully correlated across \pt and $y$ bins for the same centrality and it is reported separately from other uncertainties. For the case of the \RAA evaluated as a function of \Npart, the \TAA uncertainty is added in quadrature with other uncertainties. 
\subsection{Trigger and reconstruction efficiency uncertainty}
Several sources of systematic uncertainty were examined to assess the uncertainties of the muon efficiency determination. The statistical uncertainty of the fitted scale factors is propagated as a systematic uncertainty. The signal and background fit models used to extract the data efficiency in the T\&P method are changed to assess systematic uncertainties related to the choice of signal and background PDFs. A Chebychev polynomial is used instead of an exponential function for the background model variation, and a single Gaussian function is used instead of a weighted sum of Gaussian and CB functions for the signal mass resolution model variation. \\
For the reconstruction efficiency, the difference between the "true" muon efficiency given by the fraction of generator-level muons that are successfully reconstructed and the efficiency determined using the T\&P method in MC simulation is also assigned as a correlated systematic uncertainty. 
Similarly, an extra 3\% correlated uncertainty is added to cover the non-closure of the dimuon chain factorization. The centrality-dependent corrections have an uncertainty of $\mathcal{O}$(1\%).  These uncertainties apply to the cross sections but cancel out in the ratios of $\psi(2\mathrm{S})$ to \jpsi yields.

\subsection{Fit model uncertainty}
The uncertainty associated with the particular choice of PDFs was evaluated by varying the PDF of each component, using  ten alternative models. In each variation of the fit model, all measured quantities were recalculated and compared to the nominal fit. The root mean square of all variations was then assigned as the fit model's systematic uncertainty. The signal mass PDF was varied by replacing the $C\!B$ plus Gaussian function with a double Gaussian function, and varying parameters of the $C\!B$ model, which were originally fixed. For the signal decay time PDF, a single exponential function was changed to a double exponential function. The background mass PDFs were varied by replacing exponential functions with second-order Chebyshev polynomials in order to describe the prompt, non-prompt and double-sided background terms. Finally, the decay time resolution was varied by using a single Gaussian function in place of the double Gaussian function.

The stability of the nominal fitting procedure is quantified by comparing the yield of a randomly weighted MC simulation sample of prompt and non-prompt $J/\psi$ with the fit output of the same sample. 
The comparison shows a 1\% difference in the yield extractions and non-prompt fraction. This is assigned as an additional systematic uncertainty in the yields and non-prompt fraction value, which, however, cancels out in the $\psi$(2S) to $J/\psi$ ratio. An extra systematic uncertainty is added to the $\psi$(2S) to $J/\psi$ ratio to account for a $2\%$ bias introduced by the acceptance interpolation (see Eq.~(\ref{eq:A_def})).

%% file: Section6_Results.tex
\subsection{Prompt and non-prompt \jpsi per-event yields for \text{Pb+Pb} collisions}
\begin{figure}[htb]
  \begin{center}
  \includegraphics[width=0.49\textwidth]{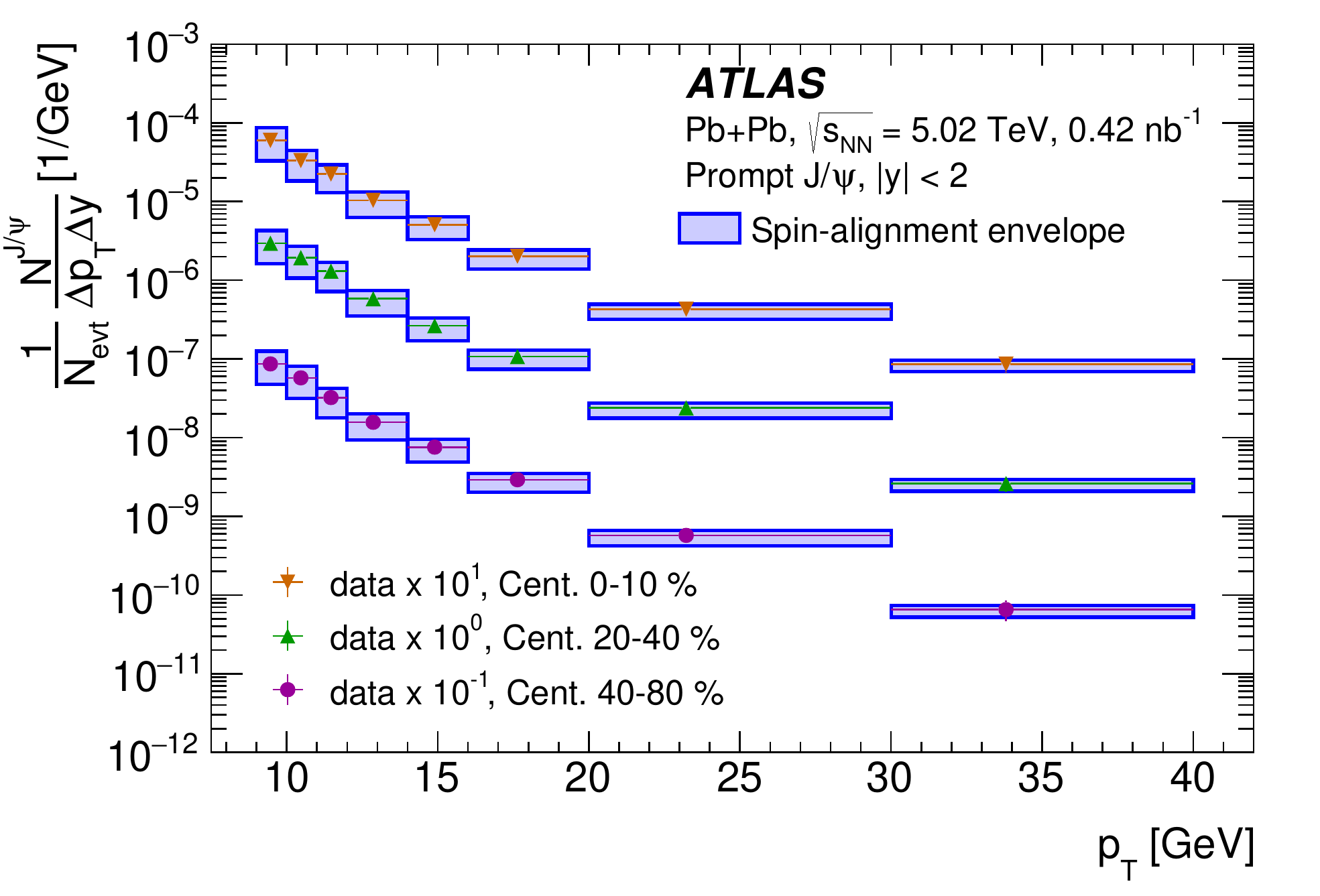}
  \includegraphics[width=0.49\textwidth]{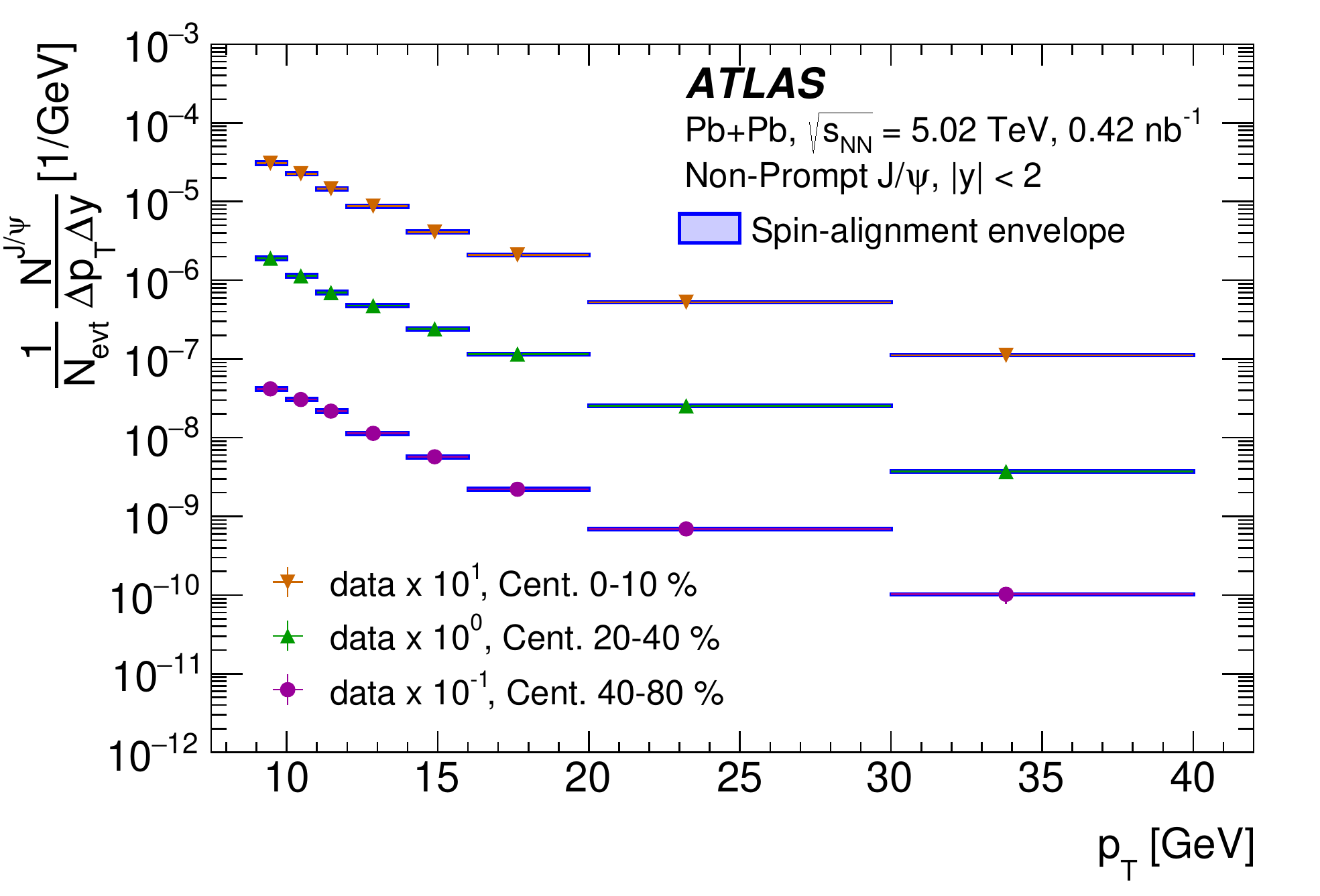}\\
  \caption{\PbPb per-event yields of prompt $\jpsi$ (left) and non-prompt $\jpsi$ (right) as a function of \pT for three different centrality slices in the rapidity range $|y| < 2$. The centroids of the \pt bins are the mean value of the transverse momentum distributions corrected for acceptance $\times$ efficiency. The vertical error bars are the combined systematic and statistical uncertainties. Overlaid is a band representing the variation of the result in various spin-alignment scenarios.}
  \label{fig:PbPb_Yields}
  \end{center}
\end{figure}

\begin{figure}[htb]
  \begin{center}
  \includegraphics[width=0.49\textwidth]{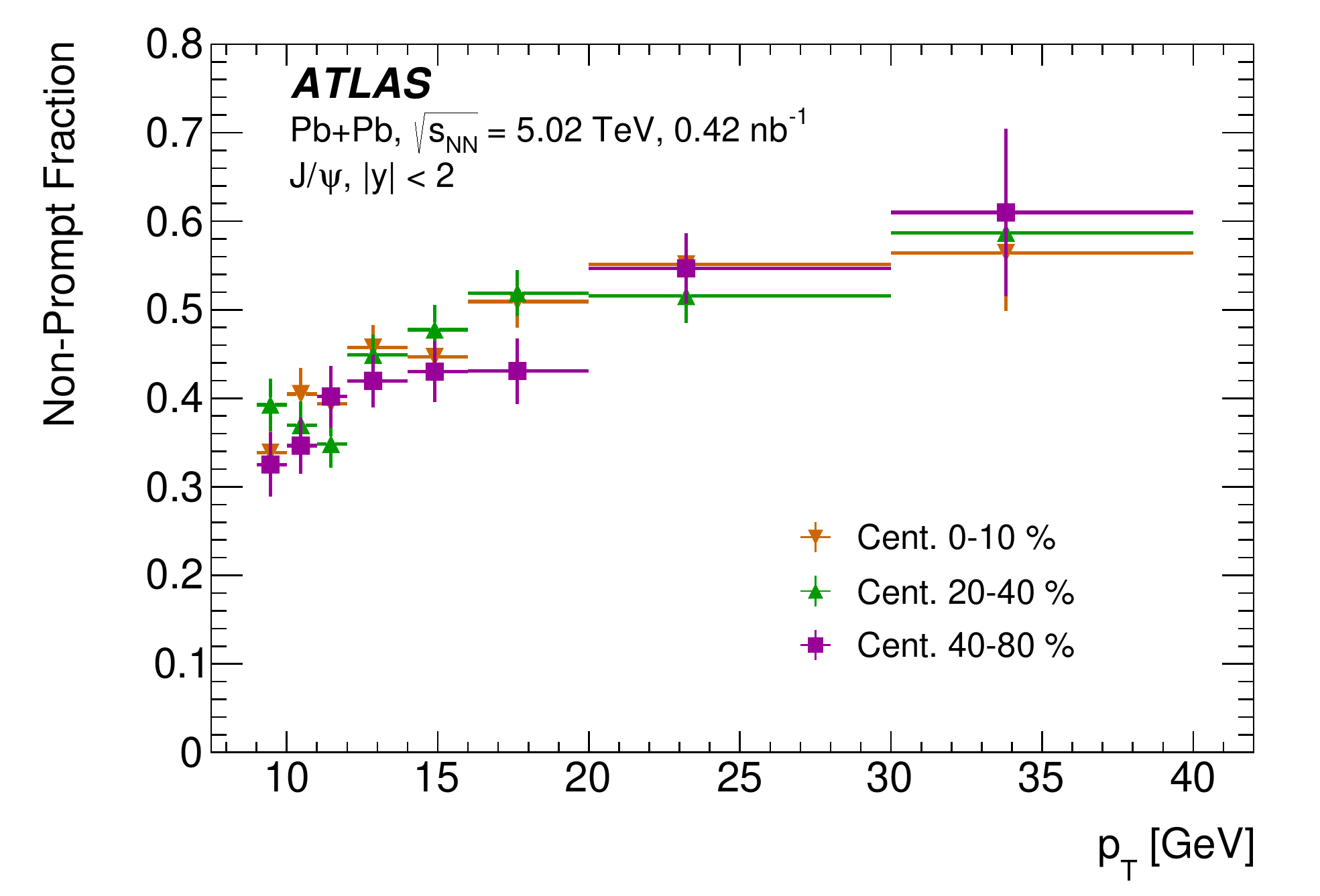}
  \includegraphics[width=0.49\textwidth]{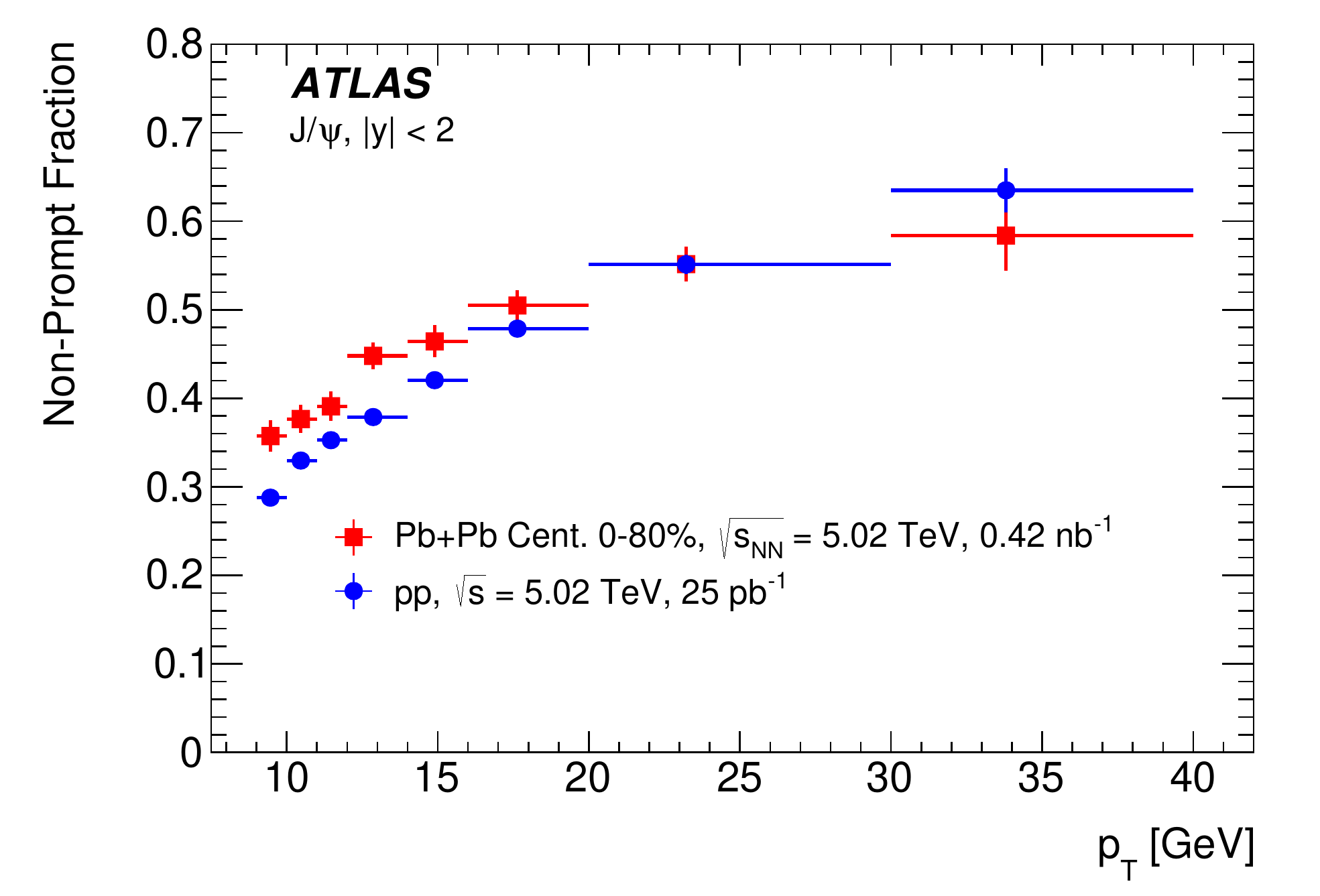}\\
  \caption{(Left) Non-prompt fraction of \jpsi production in 5.02 TeV \PbPb collision data as a function of \pT for three different centrality slices in the rapidity range $|y| < 2$. (Right) Comparison with the ATLAS 5.02 TeV \pp collision data~\cite{Qipeng:PaperpPbJpiATLAS}. The vertical error bars are the combined systematic and statistical uncertainties, dominated by the statistical uncertainty.}
  \label{fig:bfraction_PbPb}
  \end{center}
\end{figure}

The per-event yields are defined as the number of \jpsi produced per bin of \pT, $y$ and centrality intervals normalized by the width of the \pT and $y$ bin and the number of events, $N_\text{evt}$, measured in minimum-bias data for each centrality class, as defined in Eq.~(\ref{eq:prompt_perEventYield_def}). The resulting per-event yields and non-prompt fraction for \jpsi production are shown in Figures \ref{fig:PbPb_Yields} and \ref{fig:bfraction_PbPb} respectively, as a function of transverse momentum, for three centrality slices and rapidity range $|y|<2$. The vertical error bars in the $J/\psi$ per-event yields shown in Figure \ref{fig:PbPb_Yields} are the combined systematic and statistical uncertainties.
The non-prompt fraction appears to be essentially centrality-independent and to 
have a slightly different slope from that found in $pp$ collisions~\cite{Qipeng:PaperpPbJpiATLAS}.

\subsection{Nuclear modification factor, $R^{\jpsi}_\mathrm{AA}$}
\begin{figure}[htb!]
  \begin{center}
  \includegraphics[width=0.49\textwidth]{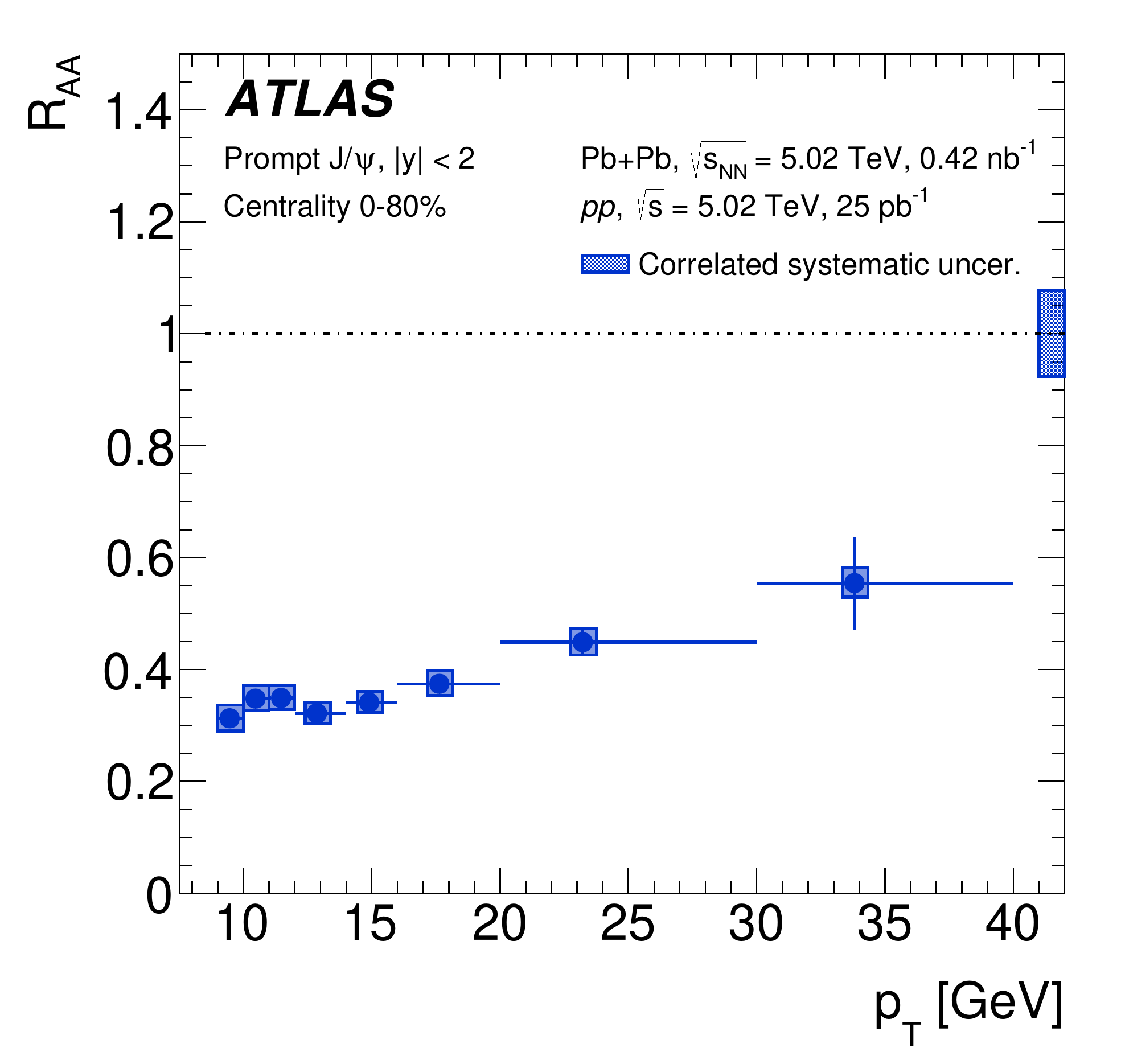}
  \includegraphics[width=0.49\textwidth]{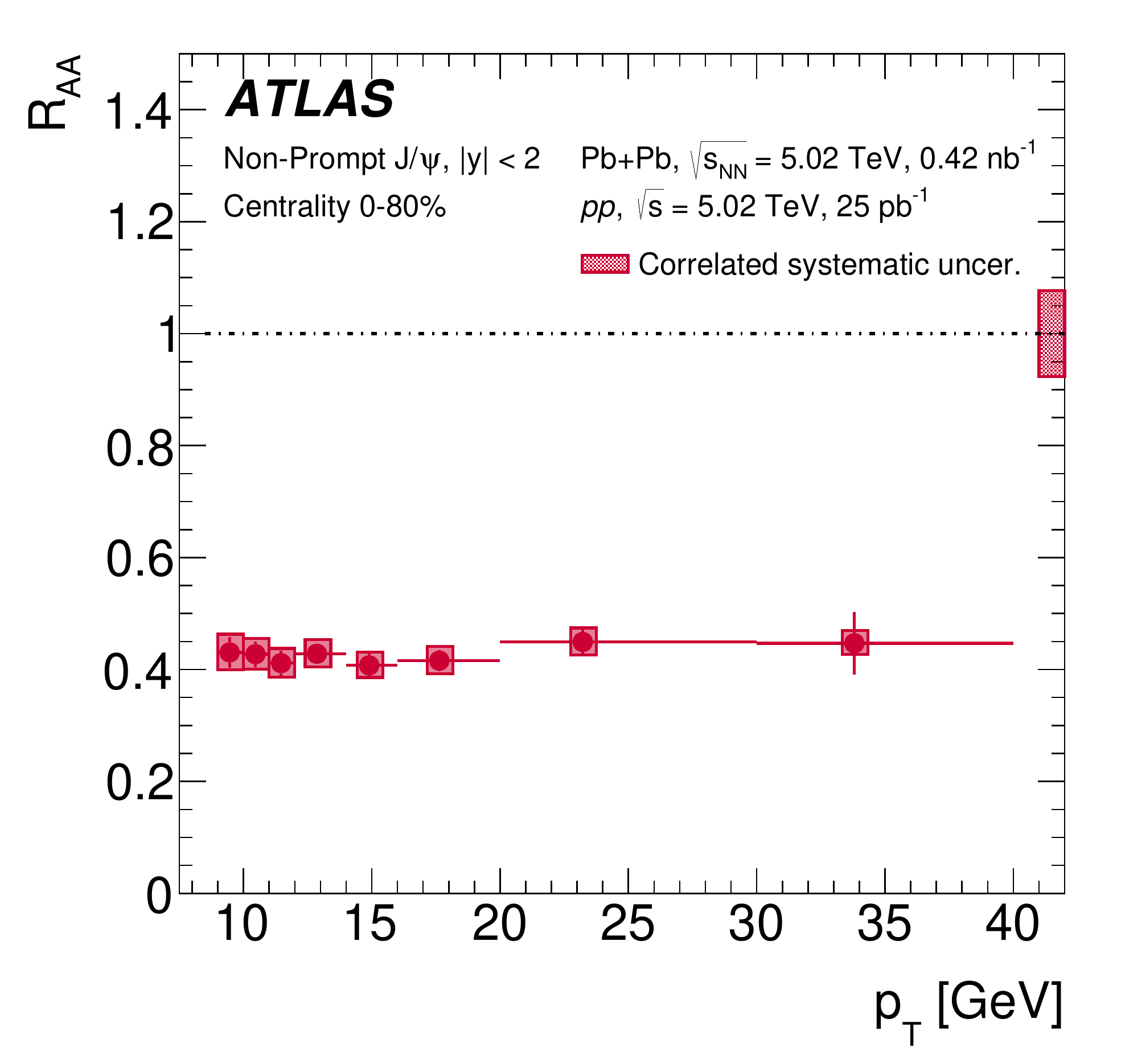}\\
  \includegraphics[width=0.49\textwidth]{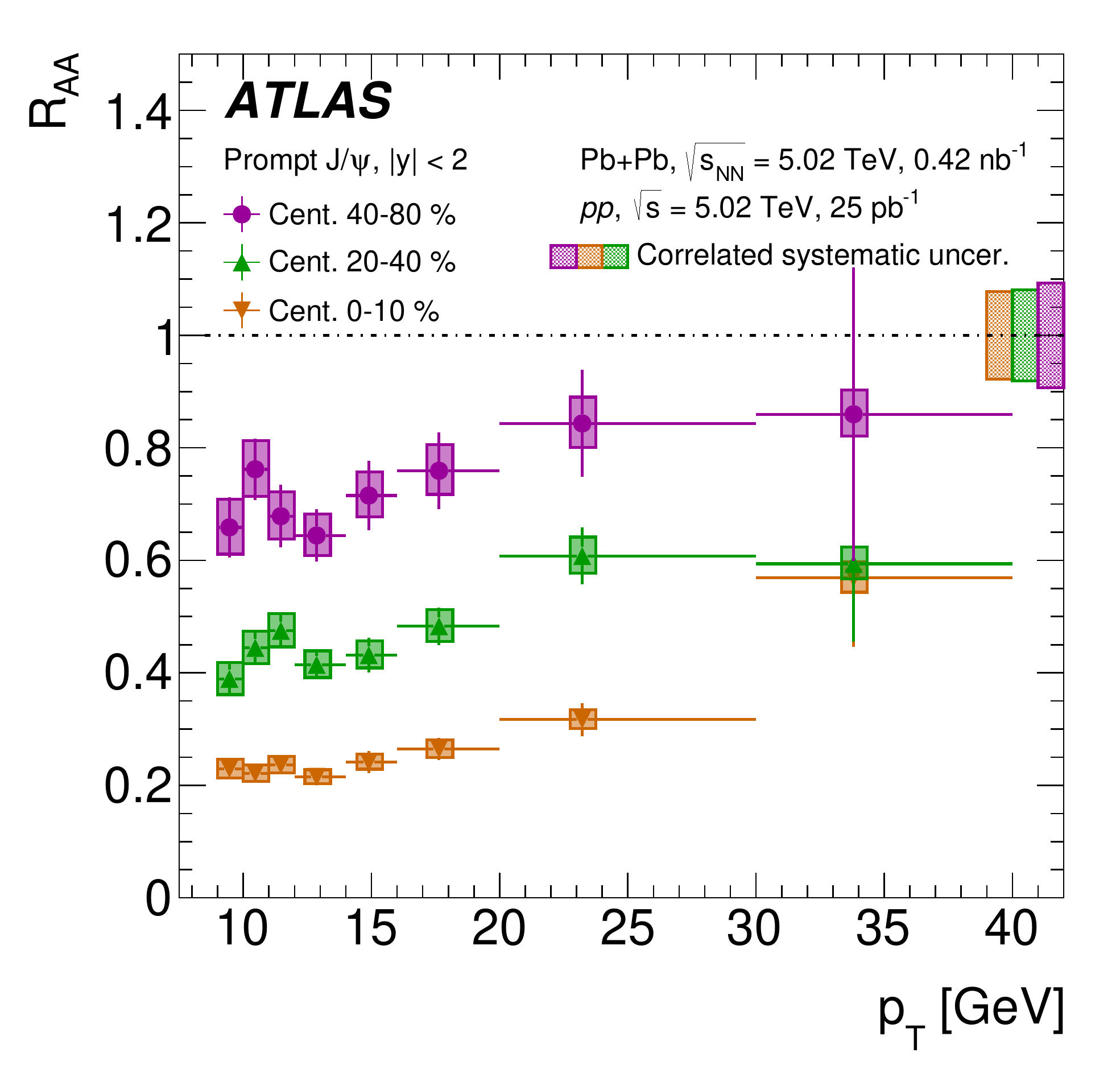}
  \includegraphics[width=0.49\textwidth]{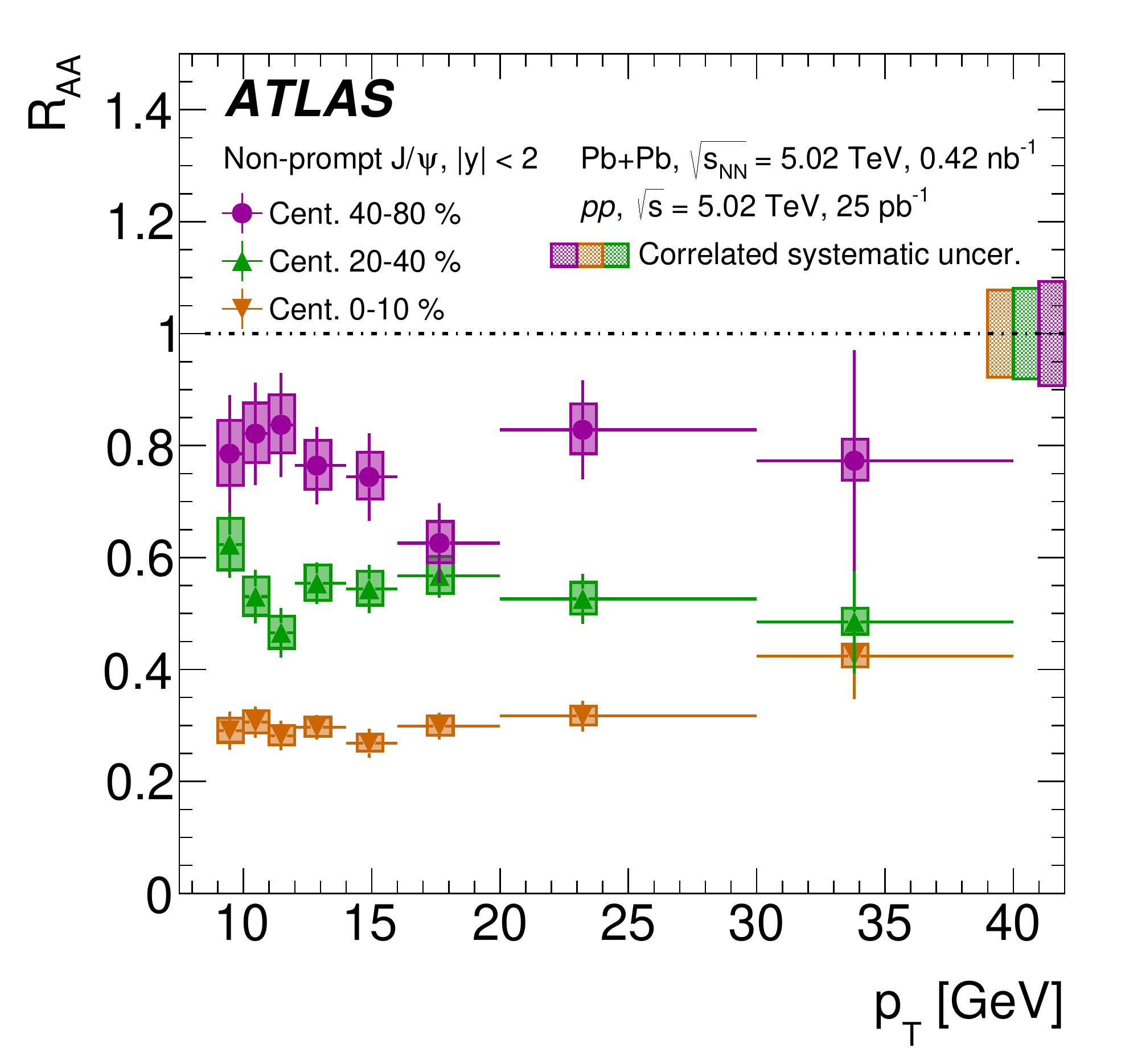}\\ 
  \caption{The nuclear modification factor as a function of \pT for the prompt $\jpsi$ (left) and non-prompt $\jpsi$ (right) for $|y|<2$, in 0--80\% centrality bin (top) and in 0--10\%, 20--40\%, and 40--80\% centrality bins (bottom). The statistical uncertainty of each point is indicated by a narrow error bar. The error box plotted with each point represents the uncorrelated systematic uncertainty, while the shaded error box at $R_{\mathrm{AA}}$=1 represents correlated scale uncertainties.} 
  \label{fig:R_AApt} 
  \end{center}
\end{figure}

\begin{figure}[htb!]
  \begin{center}
  \includegraphics[width=0.49\textwidth]{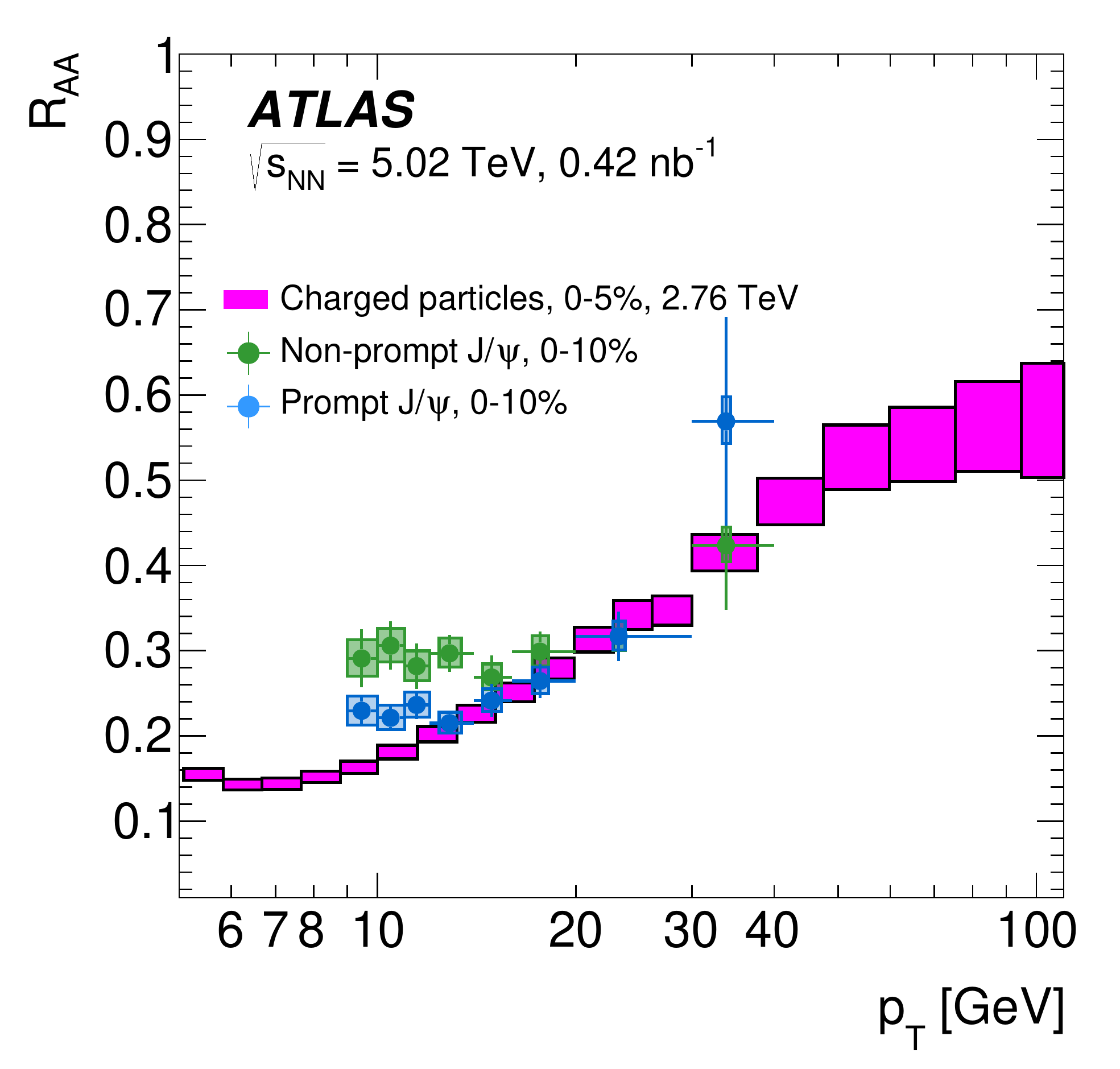}
  \includegraphics[width=0.49\textwidth]{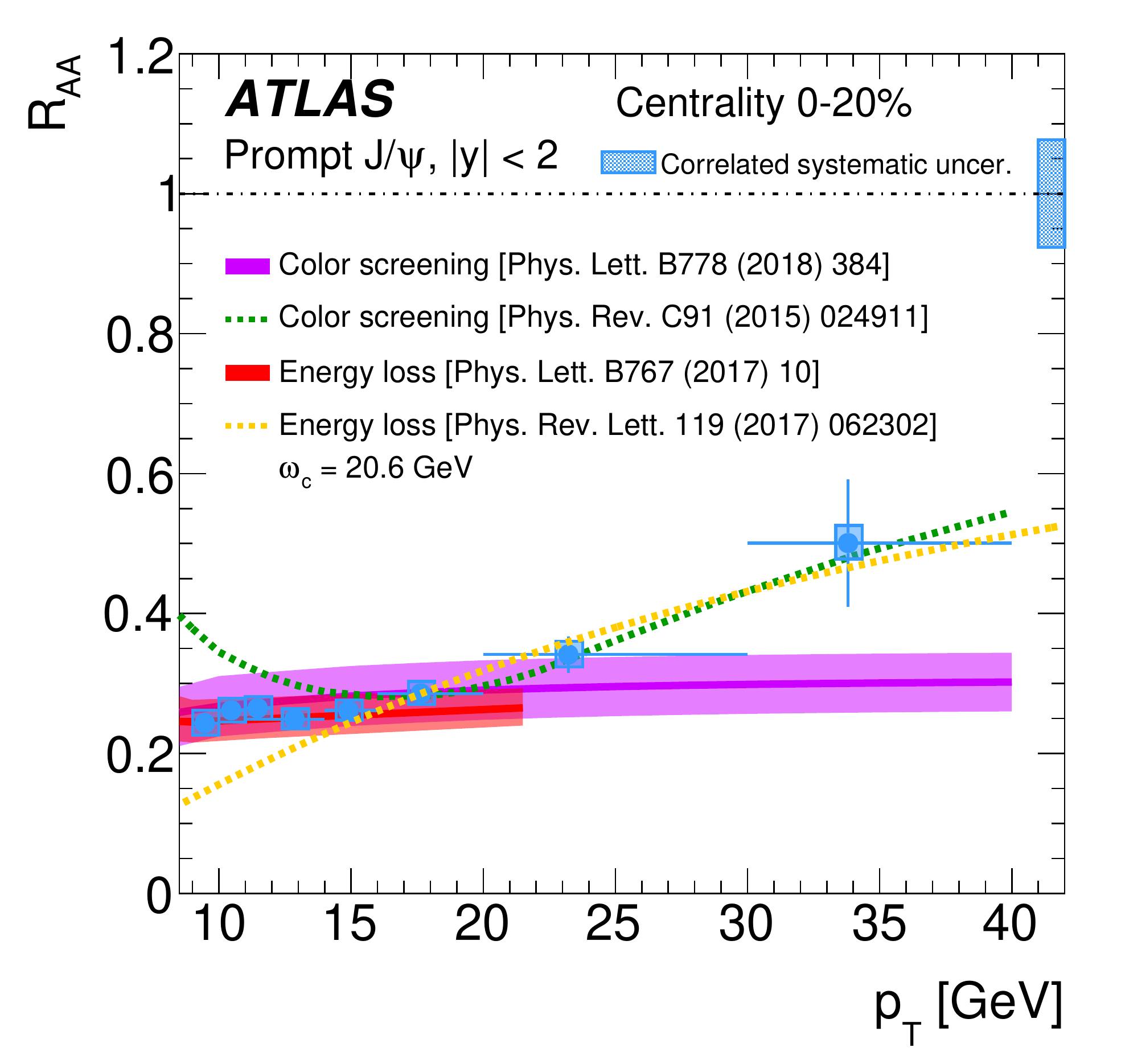}\\ 
  \caption{(Left) Comparison of prompt and non-prompt \jpsi \RAA with the \RAA of charged particles~\cite{CP_ATLAS}. (Right) Comparison of the \RAA for prompt \jpsi production with different theoretical models. The statistical uncertainty of each point is indicated by a narrow error bar. The error box plotted with each point represents the uncorrelated systematic uncertainty, while the shaded error box at $R_{\mathrm{AA}}$=1 represents correlated scale uncertainties.} 
  \label{fig:model_comparison} 
  \end{center}
\end{figure}

\begin{figure}[htb!]
  \begin{center}
  \includegraphics[width=0.49\textwidth]{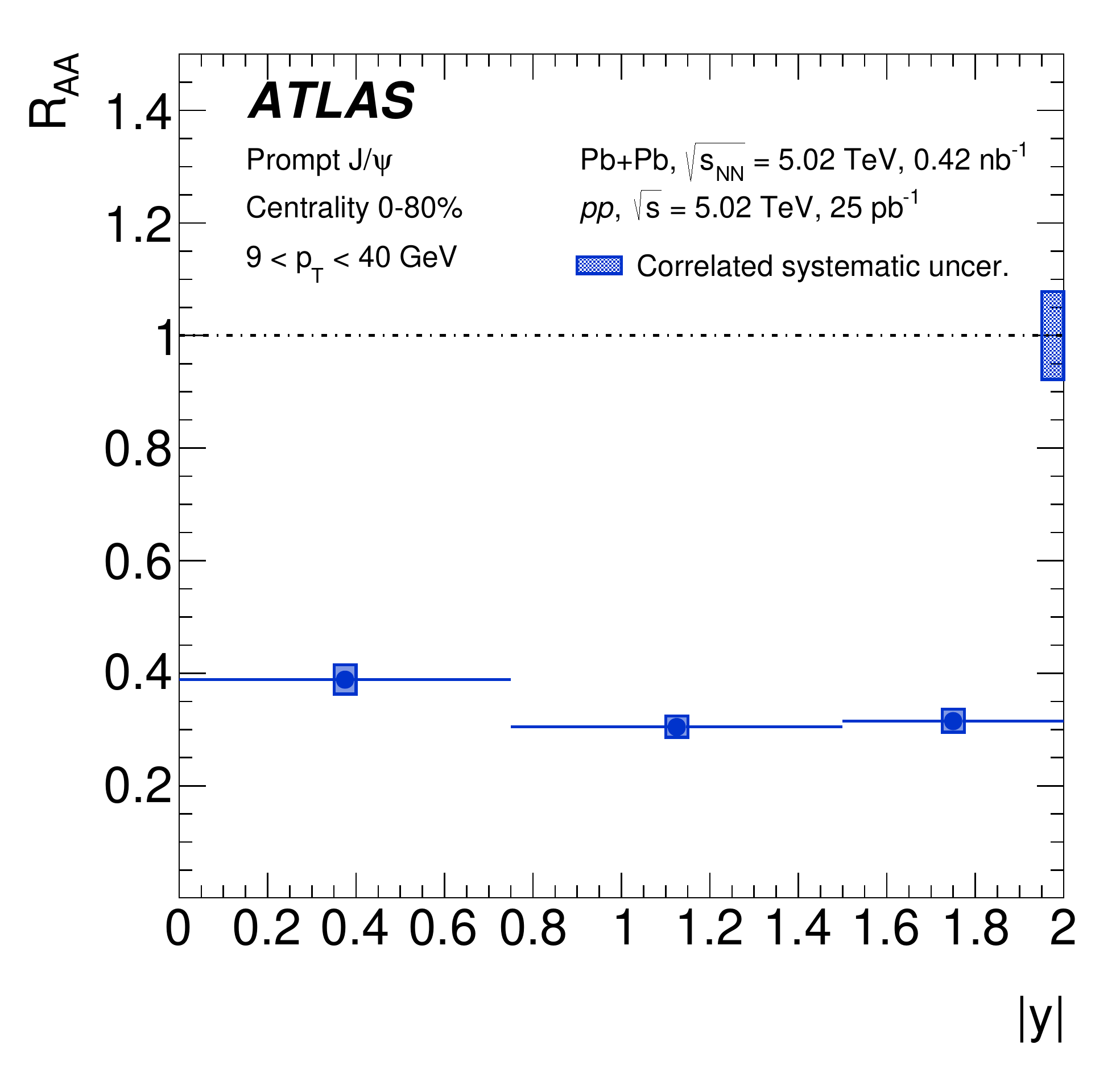}
  \includegraphics[width=0.49\textwidth]{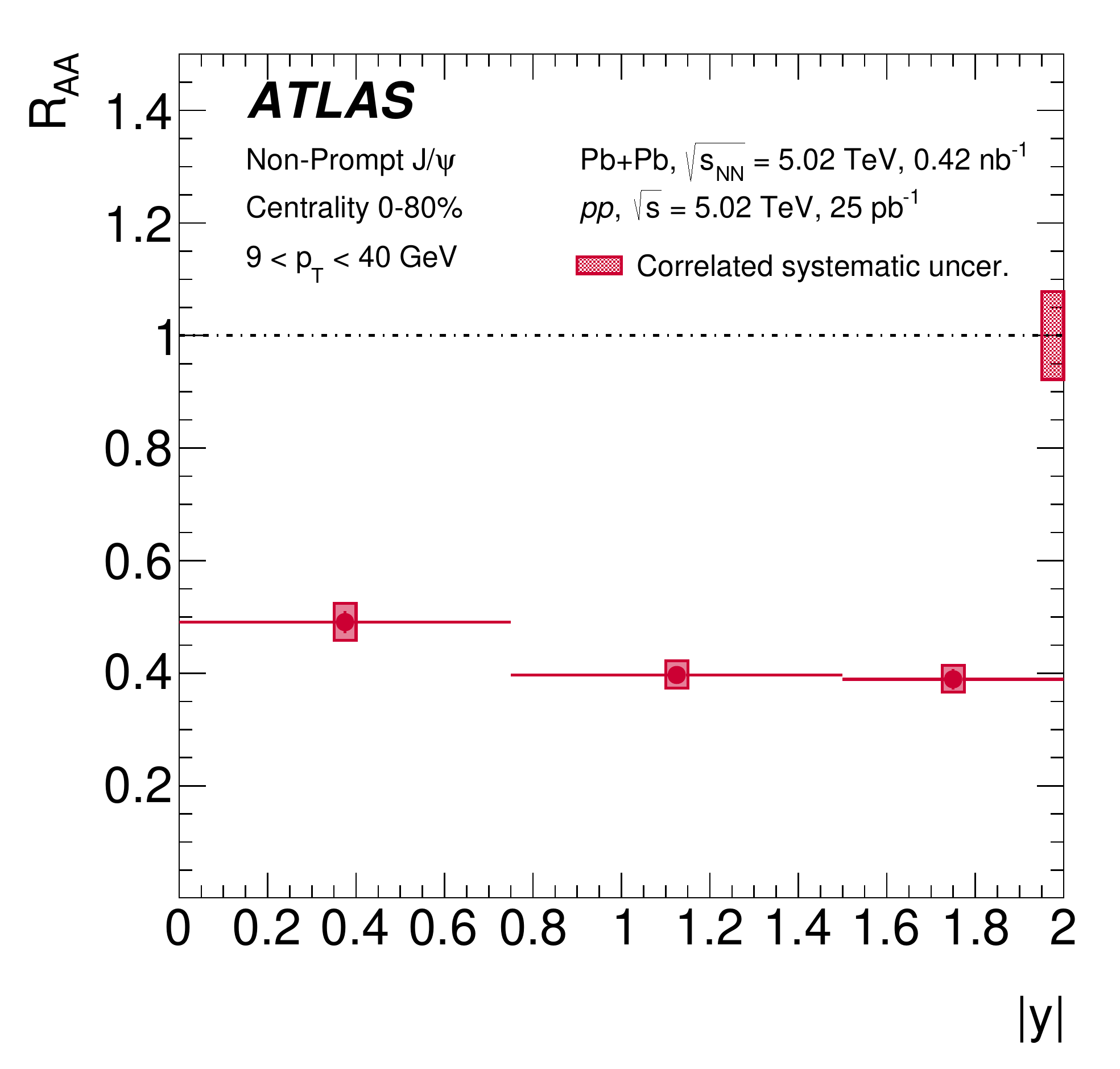}\\
  \includegraphics[width=0.49\textwidth]{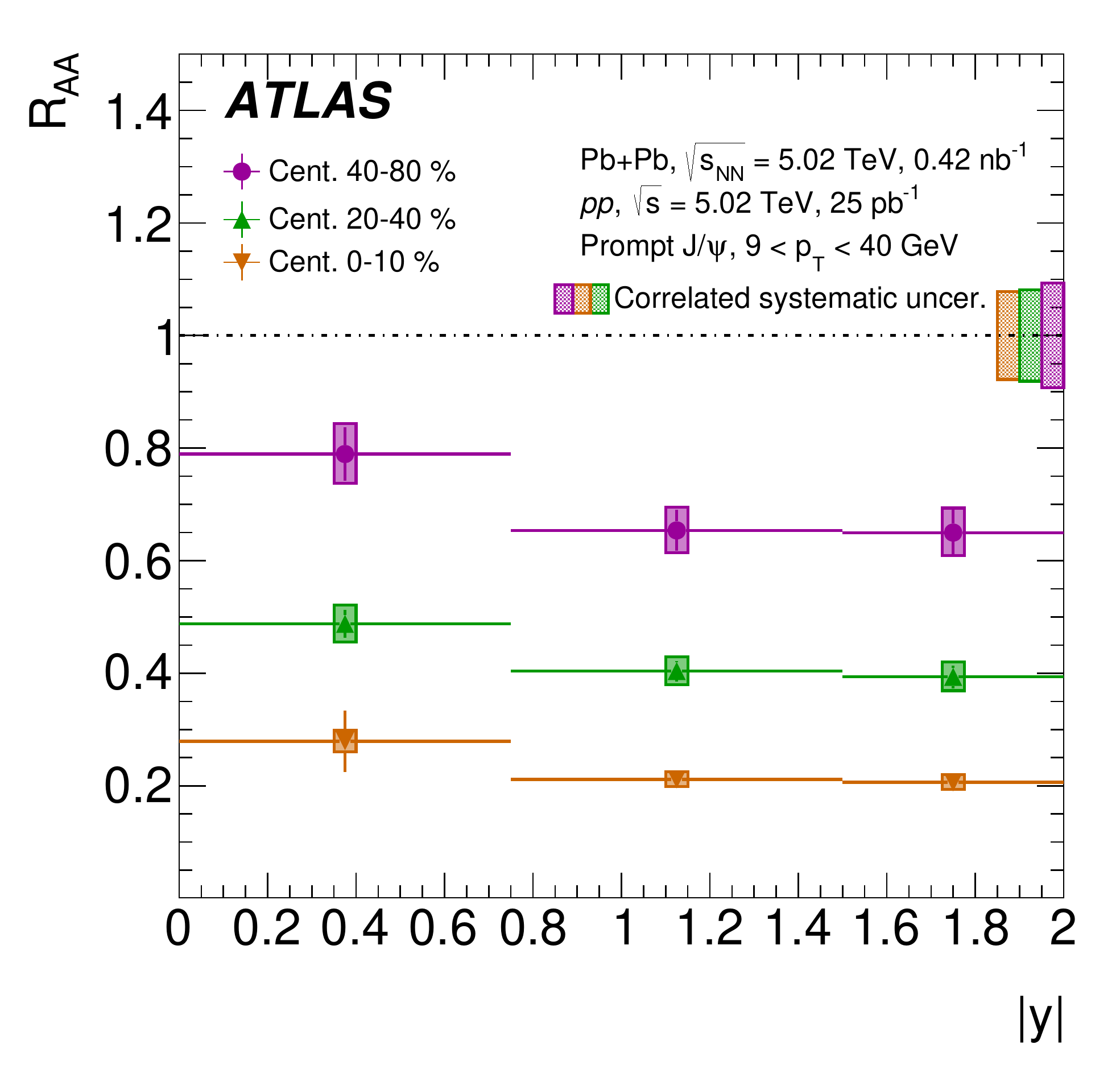}
  \includegraphics[width=0.49\textwidth]{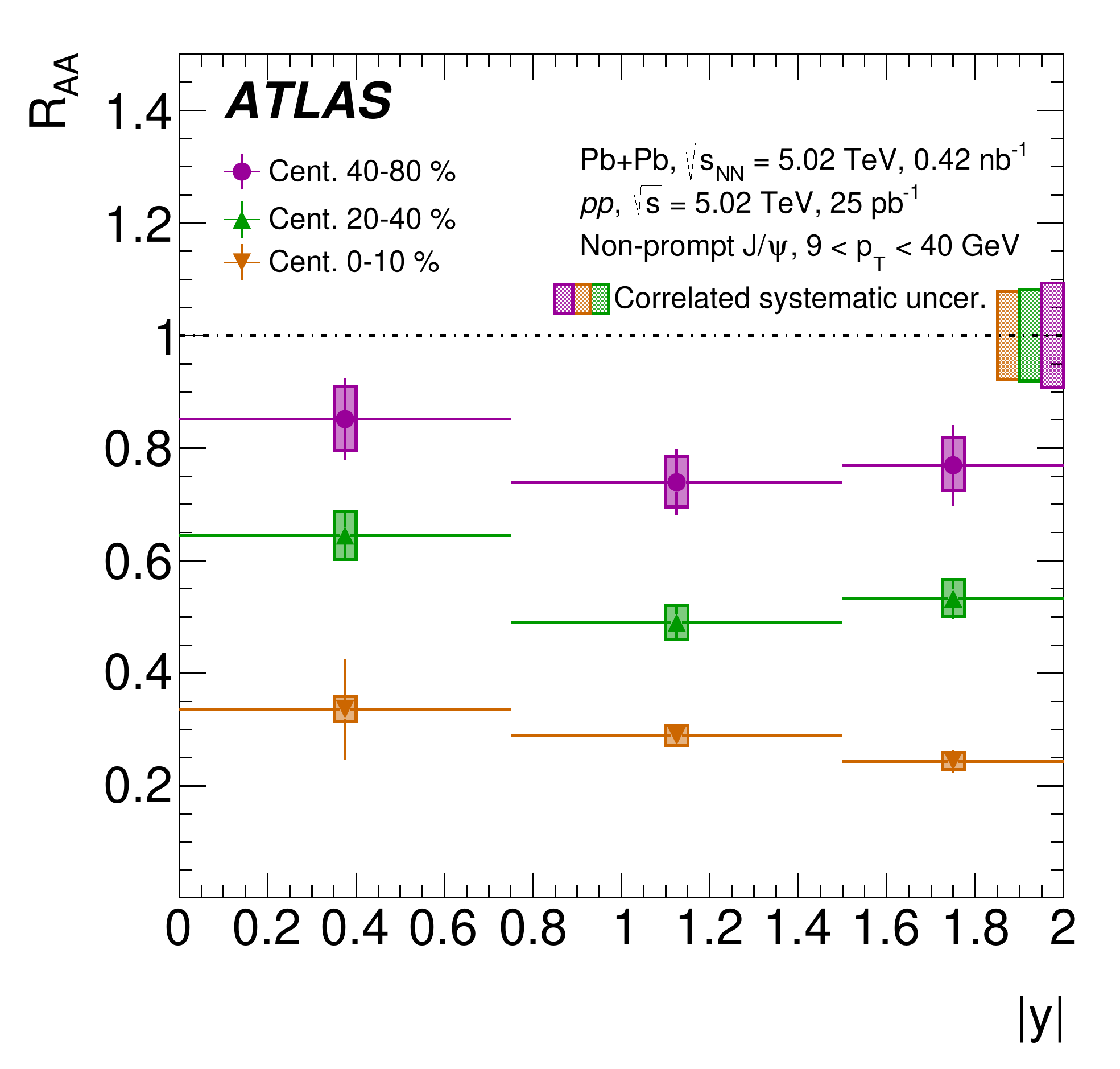}\\ 
  \caption{The nuclear modification factor as a function of rapidity for the prompt $\jpsi$ (left) and non-prompt $\jpsi$ (right) for $9 < \pT < 40\ \GeV$, in 0--80\% centrality bin (top) and in 0--10\%, 20--40\%, and 40--80\% centrality bins (bottom). The statistical uncertainty of each point is indicated by a narrow error bar. The error box plotted with each point represents the uncorrelated systematic uncertainty, while the shaded error box at $R_{\mathrm{AA}}$=1 represents correlated scale uncertainties.}
  \label{fig:R_AAy}
  \end{center}
\end{figure}

\begin{figure}[htb!]
  \begin{center}
  \includegraphics[width=0.49\textwidth]{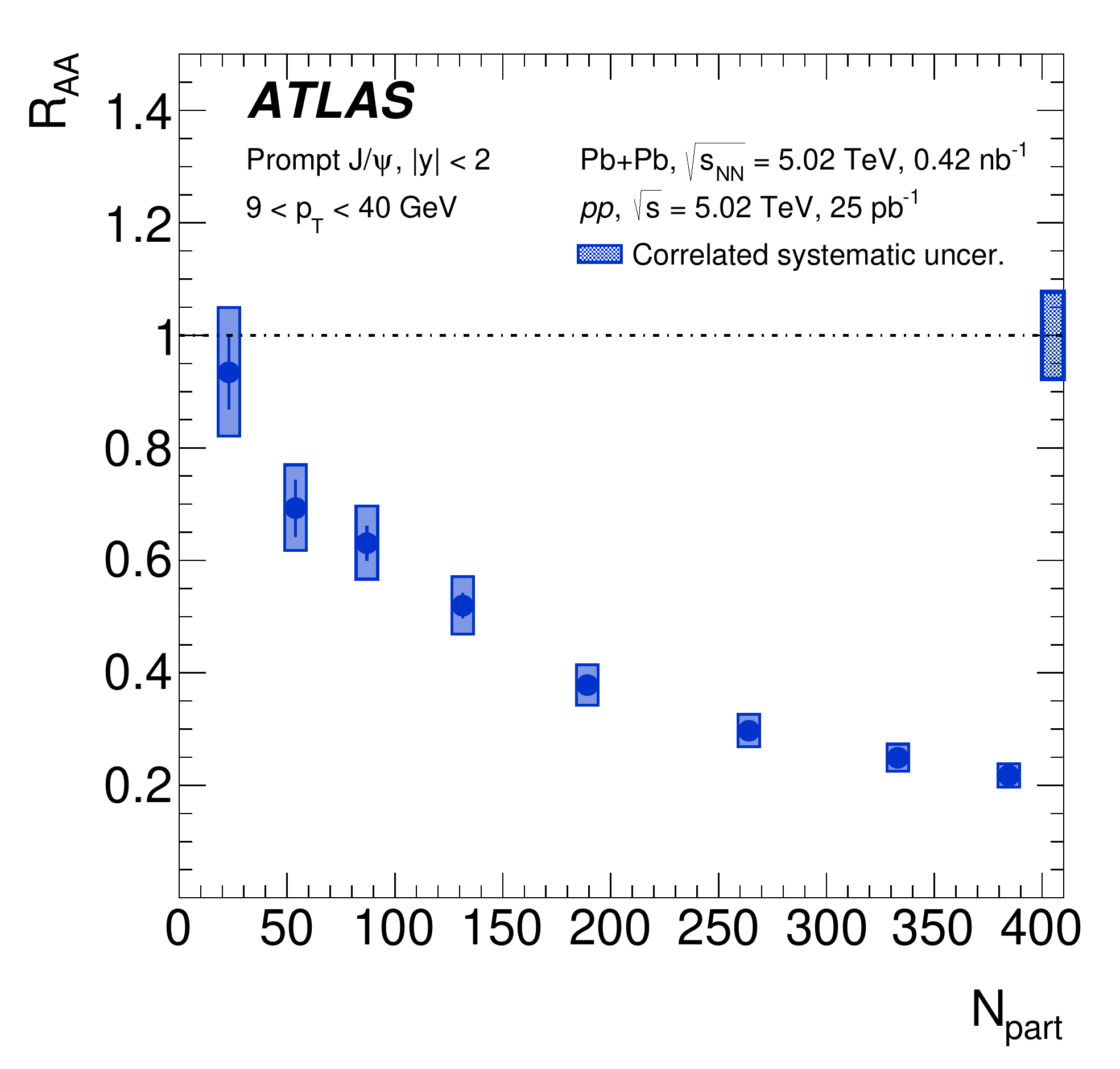}
  \includegraphics[width=0.49\textwidth]{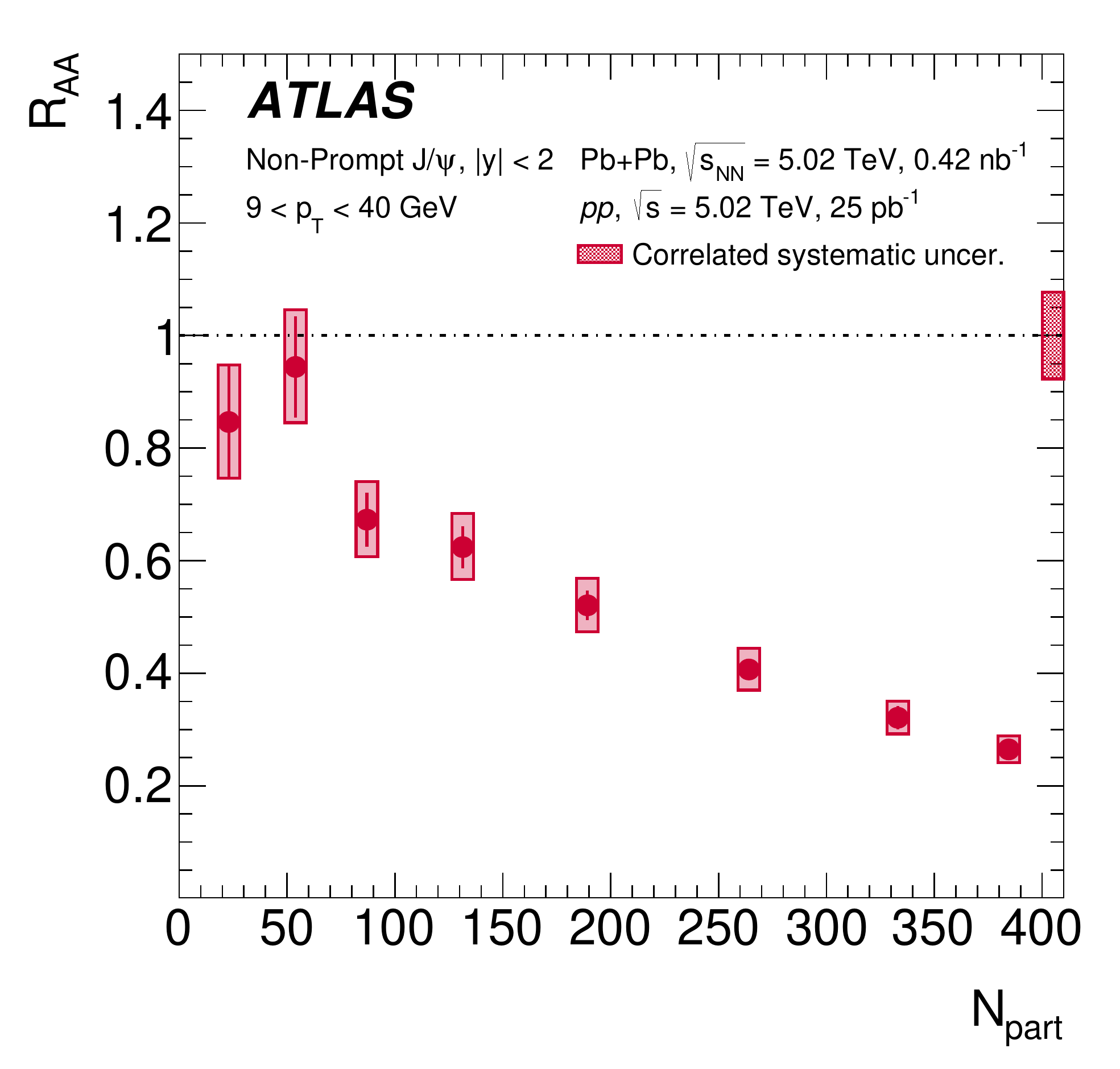}\\
  \caption{The nuclear modification factor as a function of the number of participants, $N_\mathrm{part}$, for the prompt $\jpsi$ (left) and non-prompt $\jpsi$ (right) for $9 < \pT < 40\ \GeV$ and for rapidity $|y|<2$. The statistical uncertainty of each point is indicated by a narrow error bar. The error box plotted with each point represents the uncorrelated systematic uncertainty, while the shaded error box at $R_{\mathrm{AA}}$=1 represents correlated scale uncertainties.}
  \label{fig:R_AAcent}
  \end{center}
\end{figure}

The influence of the hot dense medium on the production of the $\jpsi$ mesons is quantified by the nuclear modification factor, given in Eq.~(\ref{eq:RPbPb_def2}), which compares production of charmonium states in \PbPb collisions to the same process in \pp collisions, taking geometric factors into account. 
The results of the measurement of this observable are presented as a function of transverse momentum in Figures~\ref{fig:R_AApt}~and~\ref{fig:model_comparison}, 
rapidity in Figure~\ref{fig:R_AAy}, and centrality in Figure~\ref{fig:R_AAcent}; the last is presented as a function of the mean number of participants. 
The error box on the right-hand side of the plots located at the $R_{\mathrm{AA}}$ value of 1 indicates the correlated systematic uncertainties of the measurement, while the error boxes associated with data-points represent the uncorrelated systematic uncertainties, and the error bars indicate the statistical uncertainties. The results exhibit agreement with previous measurements performed by CMS at $\sqrt{s_\textrm{NN}}= 2.76$ and 5.02 TeV in a similar kinematic region~\cite{CMS_1,Sirunyan:2017isk}.

Figure~\ref{fig:R_AApt} shows the nuclear modification factor as a function of \pT for production of prompt and non-prompt $\jpsi$, for $|y|$ < 2.0, and for four selections of centrality.
In this figure, it can be seen that the production of $\jpsi$ is strongly suppressed in central \PbPb collisions. In the kinematic range plotted, as a function of \pT, the nuclear modification factor for both prompt and non-prompt $\jpsi$ production is seen to be in the range 0.2 < $ R_{\mathrm{AA}}$ < 1, depending on the centrality slice, having a minimum value for prompt \jpsi of 0.229 $\pm$ 0.017(stat) $\pm$ 0.016(syst) and 0.290 $\pm$ 0.034(stat) $\pm$ 0.021(syst) for the non-prompt \jpsi. For \pT~>~12~GeV, a small increase in $R_\mathrm{AA}$ with increasing \pT is observed for the first time in the prompt $\jpsi$ production, as shown in Figure~\ref{fig:model_comparison} (left), similar in shape and size to that observed for charged particles and $D$-mesons~\cite{CP_ATLAS,D_ALICE}, typically attributed to parton energy-loss processes and, for the case of charmonia, also to coherent radiation from the pre-resonant $q\bar q$ pair~\cite{spousta,orleo}. In Figure~\ref{fig:model_comparison} (right), one can see the prompt \jpsi \RAA evaluated for the 0--20\% centrality bin compared with several models, showing that the data are consistent with the colour screening and colour transparency picture~\cite{Kopeliovich:2014una,Aronson:2017ymv}, as well as parton energy-loss~\cite{spousta,orleo}. The \RAA value for non-prompt $\jpsi$ is seen to be approximately constant as a function of \pT within the uncertainties, also consistent with a parton energy-loss mechanism~\cite{cujet3}.

In Figure~\ref{fig:R_AAy}, the nuclear modification factor is presented as a function of rapidity for production of prompt and non-prompt $\jpsi$ for transverse momenta $9 < p_\mathrm{T} < 40$ GeV and for four selections of centrality.
It can be seen from the figure that the $R_\mathrm{AA}$ exhibits a modest dependence on rapidity, as expected from Ref.~\cite{Chen:2015iga}. These patterns are seen to be similar for both prompt and non-prompt $\jpsi$ production.
Figure~\ref{fig:R_AAcent} presents the nuclear modification factor as a function of centrality, expressed as the number of participants, $N_\mathrm{part}$, for production of prompt and non-prompt $\jpsi$ for $|y|$ < 2.0, and for $9 < p_\mathrm{T} < 40$~GeV.
In the kinematic range plotted, as a function of centrality, the nuclear modification factor for both prompt and non-prompt \jpsi decrease from the most peripheral bin, 60--80\%, to the most central bin, 0--5\%, with a minimum value of 0.217 $\pm$ 0.010(stat) $\pm$ 0.020(syst) for prompt and 0.264 $\pm$ 0.017(stat) $\pm$ 0.023(syst) for non-prompt.  
Suppression by a factor of about 4 or 5 for both the prompt and non-prompt $\jpsi$ mesons in central collisions is a very striking sign that the hot dense medium has a strong influence on the particle production processes. While the production of prompt $\jpsi$ mesons is found to be suppressed slightly more in the mid-centrality region, the two classes of meson production have essentially the same pattern. This is not expected because the two cases are believed to have quite different physical origins: the non-prompt production should be dominated by \textit{b}-quark processes that extend far outside the deconfined medium, whereas the prompt production happens predominantly within the medium. 
\begin{figure}[]
  \begin{center}
   \includegraphics[width=0.49\textwidth]{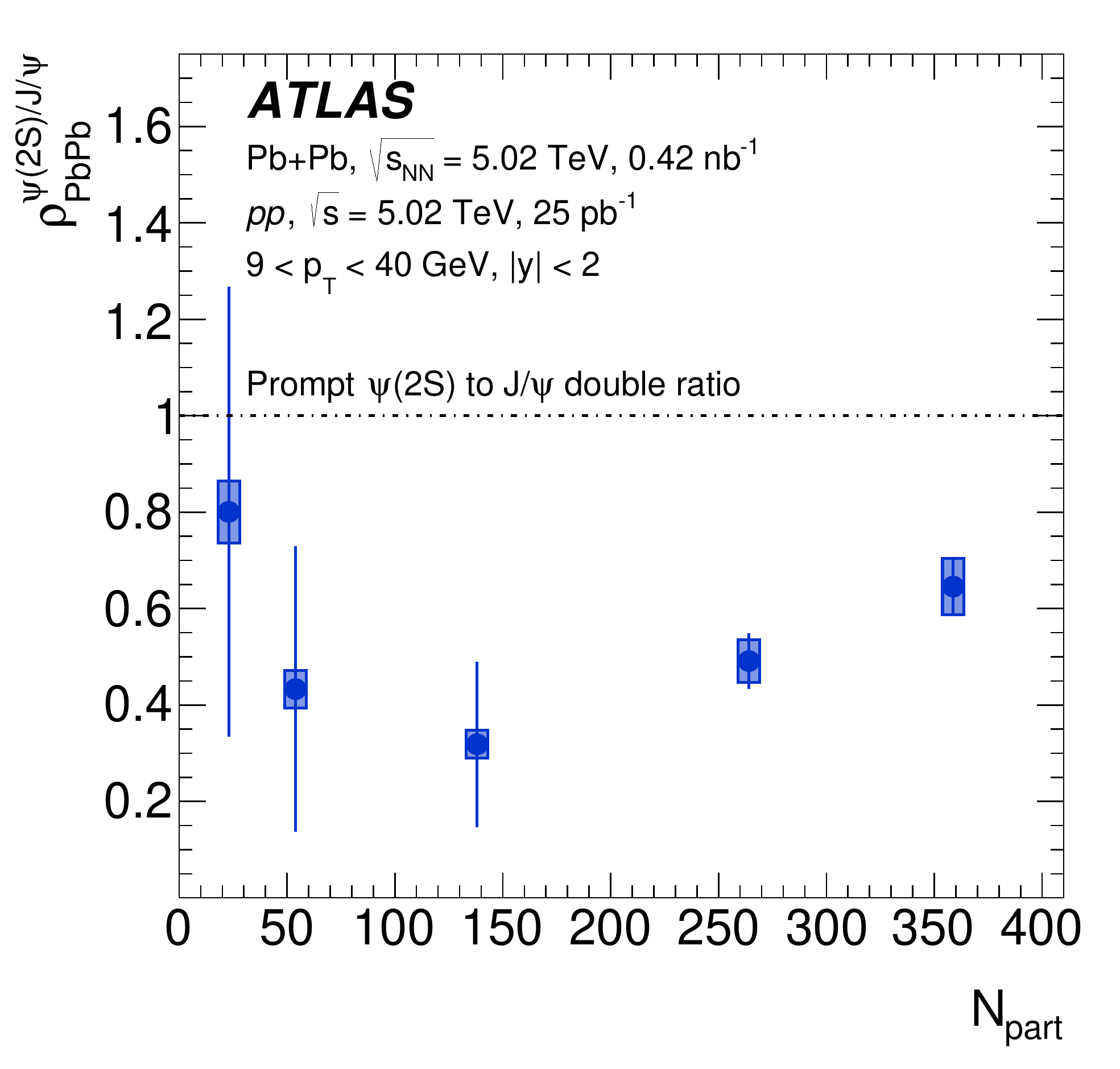}
   \includegraphics[width=0.49\textwidth]{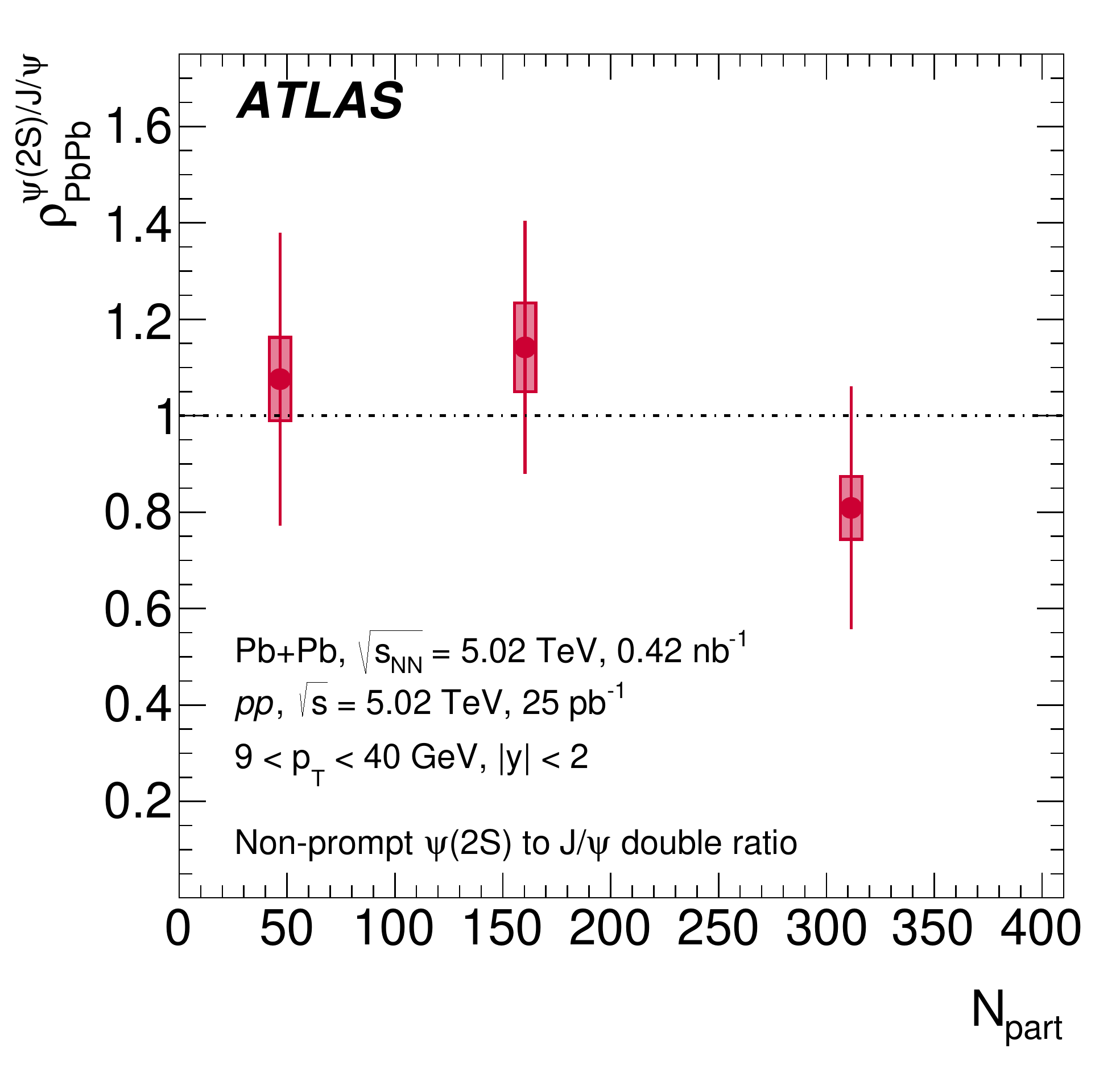}
  \caption{ $\psi(2\mathrm{S})$ to \jpsi double ratio, as a function of the number of participants, $N_\mathrm{part}$, for prompt meson production (left) and non-prompt meson production (right). The narrow error bar represents the statistical uncertainties while the error box represents the total systematic uncertainty.}
  \label{fig:R_ratio}
  \end{center}
\end{figure}
\subsection{$\psi(2\mathrm{S})$ to \jpsi yield double ratio} 
The double ratio of $\psi(2\mathrm{S})$ production to $\jpsi$ meson production, $\rho_\textrm{PbPb}^{\psi(2\mathrm{S})/\jpsi}$ is shown in Figure~\ref{fig:R_ratio}. These results  represent a measurement complementary to an earlier measurement of $\psi(2\mathrm{S})$ to \jpsi yield ratios at the same centre-of-mass energy made by the CMS Collaboration~\cite{CMS_2}. This ratio, which compares the suppression of the two mesons, can be interpreted in models in which the binding energy of the two mesons is estimated, or in which the formation mechanisms differ. In the simplest interpretation, it can also offer an important cross-check. If the non-prompt $\jpsi$ and $\psi(2\mathrm{S})$ originate from \textit{b}-quarks losing energy in the medium and hadronizing outside of the medium, then the ratio of their yields should be unity. This statement should be true for the ratio expressed as a function of any kinematic variable. By contrast, prompt $\jpsi$ and $\psi(2\mathrm{S})$ should traverse the hot and dense medium. Considering both mesons as composite systems, with potentially different formation mechanisms and different binding energies, they may respond differently to the hot dense medium.
This interpretation is supported by the results of Figure~\ref{fig:R_ratio}, which shows the ratio of $\psi(2\mathrm{S})$ to $\jpsi$ production as a function of the number of collision participants, $N_\mathrm{part}$. The ratio is consistent with unity within the experimental uncertainties for non-prompt mesons, while for prompt \jpsi the ratio is different from unity. These data support the enhanced suppression of prompt $\psi(2\mathrm{S})$ relative to $\jpsi$. This observation is consistent with the interpretation that the tightest bound quarkonium system, the $\jpsi$, survives the temperature of the hot and dense medium with a higher probability than the more loosely bound state, the $\psi(2\mathrm{S})$. It is, however, also consistent with the radiative energy-loss scenario as shown in Ref.~\cite{spousta}.
Irrespective of the underlying mechanism for the charmonium suppression, one may expect less ambiguity in the interpretation of these results since quark recombination processes, $\jpsi$s formed from uncorrelated $c\bar c$ pairs in the plasma, which are important at small $p_\mathrm{T}^{\psi(\textit{n}\mathrm{S})}$, should not play a significant role here~\cite{Thews:2000rj,BraunMunzinger:2000px,ALICE:1606.08197}. At the same time, the possible enhancement of $\psi(2\mathrm{S})$ yields seen for \Npart > 300 in the $\rho_\textrm{PbPb}^{\psi(2\mathrm{S})/\jpsi}$ distribution might be an indication of a sequential regeneration of charmonia in the most central collisions~\cite{Du:2015wha,Du:2016fam}.

%% file: Section7_Conclusion.tex
Measurements of \jpsi and $\psi(2\mathrm{S})$ production are performed in the dimuon decay channel in \PbPb collisions at $\sqrt{s_\mathrm{NN}}$ = 5.02~TeV with an integrated luminosity of  $0.42~\mathrm{nb}^{-1}$, and in $pp$ collisions at $\sqrt{s}$ = 5.02~TeV, with an integrated luminosity of $25~\ipb$ collected with the ATLAS experiment at the LHC. Results are presented for prompt and non-prompt nuclear modification factors of the \jpsi mesons, as well as the yields and non-prompt fraction in the region with transverse momentum $9 < \pt < 40\ \GeV$ and rapidity $|y| < 2.0$.

Strong suppression of prompt and non-prompt $\jpsi$ and $\psi(2\mathrm{S})$ mesons is observed in \PbPb data. The maximum suppression of prompt and non-prompt  $\jpsi$ is observed for the most central collisions. The dependence of the nuclear modification factor $R_{\mathrm{AA}}$ on centrality is approximately the same for prompt and non-prompt $\jpsi$. The prompt $\jpsi$ $R_\mathrm{AA}$, as a function of \pT, shows an increasing trend while the non-prompt $\jpsi$ $R_\mathrm{AA}$ is consistent with being constant as a function of \pT within the uncertainties.

The ratio of $\psi(2\mathrm{S})$  to $\jpsi$ meson production is measured for both the prompt and non-prompt mesons, and is shown as a function of centrality. Values consistent with unity are measured for the non-prompt mesons, while the values observed for the prompt mesons are below unity. 

%% file: acknowledgements/Acknowledgements.tex

We thank CERN for the very successful operation of the LHC, as well as the
support staff from our institutions without whom ATLAS could not be
operated efficiently.

We acknowledge the support of ANPCyT, Argentina; YerPhI, Armenia; ARC, Australia; BMWFW and FWF, Austria; ANAS, Azerbaijan; SSTC, Belarus; CNPq and FAPESP, Brazil; NSERC, NRC and CFI, Canada; CERN; CONICYT, Chile; CAS, MOST and NSFC, China; COLCIENCIAS, Colombia; MSMT CR, MPO CR and VSC CR, Czech Republic; DNRF and DNSRC, Denmark; IN2P3-CNRS, CEA-DRF/IRFU, France; SRNSFG, Georgia; BMBF, HGF, and MPG, Germany; GSRT, Greece; RGC, Hong Kong SAR, China; ISF, I-CORE and Benoziyo Center, Israel; INFN, Italy; MEXT and JSPS, Japan; CNRST, Morocco; NWO, Netherlands; RCN, Norway; MNiSW and NCN, Poland; FCT, Portugal; MNE/IFA, Romania; MES of Russia and NRC KI, Russian Federation; JINR; MESTD, Serbia; MSSR, Slovakia; ARRS and MIZ\v{S}, Slovenia; DST/NRF, South Africa; MINECO, Spain; SRC and Wallenberg Foundation, Sweden; SERI, SNSF and Cantons of Bern and Geneva, Switzerland; MOST, Taiwan; TAEK, Turkey; STFC, United Kingdom; DOE and NSF, United States of America. In addition, individual groups and members have received support from BCKDF, the Canada Council, CANARIE, CRC, Compute Canada, FQRNT, and the Ontario Innovation Trust, Canada; EPLANET, ERC, ERDF, FP7, Horizon 2020 and Marie Sk{\l}odowska-Curie Actions, European Union; Investissements d'Avenir Labex and Idex, ANR, R{\'e}gion Auvergne and Fondation Partager le Savoir, France; DFG and AvH Foundation, Germany; Herakleitos, Thales and Aristeia programmes co-financed by EU-ESF and the Greek NSRF; BSF, GIF and Minerva, Israel; BRF, Norway; CERCA Programme Generalitat de Catalunya, Generalitat Valenciana, Spain; the Royal Society and Leverhulme Trust, United Kingdom.

The crucial computing support from all WLCG partners is acknowledged gratefully, in particular from CERN, the ATLAS Tier-1 facilities at TRIUMF (Canada), NDGF (Denmark, Norway, Sweden), CC-IN2P3 (France), KIT/GridKA (Germany), INFN-CNAF (Italy), NL-T1 (Netherlands), PIC (Spain), ASGC (Taiwan), RAL (UK) and BNL (USA), the Tier-2 facilities worldwide and large non-WLCG resource providers. Major contributors of computing resources are listed in Ref.~\cite{ATL-GEN-PUB-2016-002}.

%% file: atlas_authlist.tex
\begin{flushleft}
{\Large The ATLAS Collaboration}

\bigskip

M.~Aaboud$^\textrm{\scriptsize 137d}$,
G.~Aad$^\textrm{\scriptsize 88}$,
B.~Abbott$^\textrm{\scriptsize 115}$,
O.~Abdinov$^\textrm{\scriptsize 12}$$^{,*}$,
B.~Abeloos$^\textrm{\scriptsize 119}$,
S.H.~Abidi$^\textrm{\scriptsize 161}$,
O.S.~AbouZeid$^\textrm{\scriptsize 139}$,
N.L.~Abraham$^\textrm{\scriptsize 151}$,
H.~Abramowicz$^\textrm{\scriptsize 155}$,
H.~Abreu$^\textrm{\scriptsize 154}$
% ...
\bigskip
\\
% ...
$^{12}$ Institute of Physics, Azerbaijan Academy of Sciences, Baku, Azerbaijan\\
% ...
$^{88}$ CPPM, Aix-Marseille Universit{\'e} and CNRS/IN2P3, Marseille, France\\
% ...
$^{115}$ Homer L. Dodge Department of Physics and Astronomy, University of Oklahoma, Norman OK, United States of America\\
% ...
$^{119}$ LAL, Universit{\'e} Paris-Sud, CNRS/IN2P3, Universit{\'e} Paris-Saclay, Orsay, France\\
% ...
$^{137}$ $^{(a)}$ Facult{\'e} des Sciences Ain Chock, R{\'e}seau Universitaire de Physique des Hautes Energies - Universit{\'e} Hassan II, Casablanca; $^{(b)}$ Centre National de l'Energie des Sciences Techniques Nucleaires, Rabat; $^{(c)}$ Facult{\'e} des Sciences Semlalia, Universit{\'e} Cadi Ayyad, LPHEA-Marrakech; $^{(d)}$ Facult{\'e} des Sciences, Universit{\'e} Mohamed Premier and LPTPM, Oujda; $^{(e)}$ Facult{\'e} des sciences, Universit{\'e} Mohammed V, Rabat, Morocco\\
% ...
$^{139}$ Santa Cruz Institute for Particle Physics, University of California Santa Cruz, Santa Cruz CA, United States of America\\
% ...
$^{151}$ Department of Physics and Astronomy, University of Sussex, Brighton, United Kingdom\\
% ...
$^{154}$ Department of Physics, Technion: Israel Institute of Technology, Haifa, Israel\\
$^{155}$ Raymond and Beverly Sackler School of Physics and Astronomy, Tel Aviv University, Tel Aviv, Israel\\
% ...
$^{161}$ Department of Physics, University of Toronto, Toronto ON, Canada\\
% ...
$^{a}$ Also at Department of Physics, King's College London, London, United Kingdom\\
$^{b}$ Also at Institute of Physics, Azerbaijan Academy of Sciences, Baku, Azerbaijan\\
$^{c}$ Also at Novosibirsk State University, Novosibirsk, Russia\\
$^{d}$ Also at TRIUMF, Vancouver BC, Canada\\
$^{e}$ Also at Department of Physics and Astronomy, University of Louisville, Louisville, KY, United States of America\\
% ...
$^{*}$ Deceased
\end{flushleft}
